\documentclass[]{pasj01}
\Received{\today}%{yyyy/mm/dd}
\Accepted{\today}%{yyyy/mm/dd}
\Published{\today}%{yyyy/mm/dd}

\jyear{2023}

\usepackage{url}
 
\newcommand{\Ni}{\ensuremath{^{56}\mathrm{Ni}}}

\newcommand{\Msun}{\ensuremath{\mathrm{M}_\odot}}
\newcommand{\Rsun}{\ensuremath{\mathrm{R}_\odot}}
\newcommand{\Msunpyr}{\ensuremath{\Msun~\mathrm{yr^{-1}}}}
\newcommand{\kmps}{\ensuremath{\mathrm{km~s^{-1}}}}

\graphicspath{{./}{figures/}}
 
\begin{document} 

\title{ 
%\LETTERLABEL %%% <-- uncomment for LETTER article  
%\REVIEWLABEL %%% <-- uncomment for REVIEW article  
Synthetic red supergiant explosion model grid for systematic characterization of Type~II supernovae
}

%%% begin:list of authors
% Do NOT capitalize all letters in "textsc".
\author{Takashi J. \textsc{Moriya}\altaffilmark{1,2}%
%\thanks{Example: Present Address is xxxxxxxxxx}
}
\altaffiltext{1}{National Astronomical Observatory of Japan, National Institutes of Natural Sciences, 2-21-1 Osawa, Mitaka, Tokyo 181-8588, Japan}
\altaffiltext{2}{School of Physics and Astronomy, Faculty of Science, Monash University, Clayton, Victoria 3800, Australia}
\email{takashi.moriya@nao.ac.jp}

\author{Bhagya M. \textsc{Subrayan},\altaffilmark{3}}
\altaffiltext{3}{Department of Physics and Astronomy, Purdue University, 525 Northwestern Ave, West Lafayette, IN 47907, USA}
%\email{bbbbb@xxx.xxx.xx.xx}

\author{Dan \textsc{Milisavljevic}\altaffilmark{3,4}}
\altaffiltext{4}{Integrative Data Science Initiative, Purdue University, West Lafayette, IN 47907, USA}
%\email{ccccc@xxx.xxx.xx.xx}

\author{Sergei I. \textsc{Blinnikov}\altaffilmark{5,6}}
\altaffiltext{5}{NRC Kurchatov Institute, 123182 Moscow, Russia}
\altaffiltext{6}{Dukhov Automatics Research Institute (VNIIA), 127055 Moscow, Russia}
%\email{ccccc@xxx.xxx.xx.xx}

%%% end:list of authors

%% `\KeyWords{}' always has to be placed before ``\maketitle'' 
%%  List of Key Words:  https://academic.oup.com/pasj/pages/Pasj_Keywords 

\maketitle

\begin{abstract}
A new model grid containing 228,016 synthetic red supergiant explosions (Type~II supernovae) is introduced. Time evolution of spectral energy distributions from 1~\AA\ to 50,000~\AA\ (100 frequency bins in a log scale) is computed at each time step up to 500~days after explosion in each model. We provide light curves for the filters of  the Vera C. Rubin Observatory's Legacy Survey of Space and Time (LSST), Zwicky Transient Facility (ZTF), Sloan Digital Sky Servey (SDSS), and the Neil Gehrels Swift Observatory, but light curves for any photometric filters can be constructed by convolving any filter response functions to the synthetic spectral energy distributions. We also provide bolometric light curves and photosphere information such as photospheric velocity evolution. The parameter space covered by the model grid is five progenitor masses ($10$, $12$, $14$, $16$, and $18~\Msun$ at the zero-age main sequence, solar metallicity), ten explosion energies ($0.5$, $1.0$, $1.5$, $2.0$, $2.5$, $3.0$, $3.5$, $4.0$, $4.5$, and $5.0\times 10^{51}~\mathrm{erg}$), nine \Ni\ masses ($0.001$, $0.01$, $0.02$, $0.04$, $0.06$, $0.08$, $0.1$, $0.2$, and $0.3~\Msun$), nine mass-loss rates ($10^{-5.0}$, $10^{-4.5}$, $10^{-4.0}$, $10^{-3.5}$, $10^{-3.0}$, $10^{-2.5}$, $10^{-2.0}$, $10^{-1.5}$, and $10^{-1.0}~\Msunpyr$ with a wind velocity of 10~\kmps), six circumstellar matter radii ($1,2,4,6,8,$ and $10\times 10^{14}~
\mathrm{cm}$), and ten circumstellar structures ($\beta=0.5,1.0,1.5,2.0,2.5,3.0,3.5,4.0,4.5,$ and $5.0$). \Ni\ is assumed to be uniformly mixed up to the half mass of a hydrogen-rich envelope. This model grid can be a base for rapid characterizations of Type~II supernovae with sparse photometric sampling expected in LSST through a Bayesian approach, for example. The model grid is available at \url{https://doi.org/10.5061/dryad.pnvx0k6sj}.
\end{abstract}

%\linenumbers

\section{Introduction}
The number of supernova (SN) discovery has increased dramatically in the last decade thanks to many large-scale optical transient surveys (e.g., \cite{nicholl2021}). The increase of SN discovery led to discoveries of rare SNe such as superluminous SNe (e.g., \cite{quimby2011}) and ultra-stripped SNe (e.g., \cite{de2018}). At the same time, the number of commonly observed SNe has also increased explosively. For example, hundreds of Type~II SNe have been reported in a year in the last few years\footnote{\url{https://www.wis-tns.org/}}. We may even have $10^5$ Type~II SNe per year in the coming era of the Vera C. Rubin Observatory's LSST \citep{lsst2009}, although the LSST survey strategy has not been fixed yet. 

Having a large number of SNe allows us to study their statistical properties. Especially, statistical studies of Type~II SNe provide us the standard nature of core-collapse SN explosions because they are the most common explosions of massive stars \citep{li2011,shivvers2017,perley2020}. Light-curve modeling of a large number of Type~II SNe has been used to estimate the general properties of Type~II SNe such as their progenitor mass, explosion energy, and \Ni\ mass (e.g., \cite{utrobin2013,nagy2014,pumo2017,morozova2018,martinez2022ii}). The estimated properties have been compared to predictions from SN explosion simulations so that the standard SN explosion mechanism can be revealed (e.g., \cite{ugliano2012,nakamura2015,muller2016,sukhbold2016,burrows2021}). Having a large number of Type~II SNe will also be beneficial for the use of Type~II SNe as a distance indicator (e.g., \cite{dejaeger2020,dejaeger2017,gall2018,nadyozhin2003,hamuy2002}).

The systematic light-curve modeling of a large number of Type~II SNe has been mostly performed by using bolometric light curves because of the difficulty in modeling multi-frequency light curves. Usually, bolometric light curves are constructed from observations, and then they are compared with theoretical models for parameter estimations (e.g., \cite{martinez2022i,martinez2022ii,martinez2022iii}). However, this kind of studies are only possible for well-observed Type~II SNe with which bolometric light curves can be estimated. Unfortunately, ongoing large-scale transient surveys such as ZTF \citep{bellm2019} have only a few bands to cover a wide field frequently, and it is difficult to construct bolometric light curves based only on photometry data provided by ZTF. The LSST survey may also not provide densely sampled photometric data covering a wide wavelength range required to estimate bolometric luminosity. In order to make full use of a large number of Type~II SNe discovered by ZTF, LSST, and other large-scale transient surveys, it is ideal to make a direct comparison between theoretical models and observations in observed bands without constructing bolometric light curves.

Light-curve modeling of Type~II SNe in multi-frequencies has been performed, but they are limited to well-observed Type~II SNe (e.g., \cite{baklanov2005,tominaga2009,hiller2019,kozyreva2022}). Multi-frequency light-curve modeling takes time and it has been difficult to perform such modeling systematically for a large number of Type~II SNe. However, \citet{forster2018} demonstrated that a systematic Bayesian parameter estimation for a large number of Type~II SNe with two-band observations is feasible. They adopted a large grid of pre-computed multi-frequency light-curve models by \citet{moriya2018}, and used it to estimate the properties of circumstellar matter (CSM) around Type~II SNe. A drawback of this approach is that pre-computed multi-frequency light-curve models need to cover a wide parameter range. Because \citet{forster2018} focused on the rising part of Type~II SNe, the adopted models had a fixed amount of \Ni\ that does not affect early light curves of Type~II SNe. Still, hundreds of pre-computed models were required for the Bayesian parameter estimation to reveal that most Type~II SNe are affected by CSM interaction in early phases (see also \cite{morozova2017}).

In this paper, we introduce a new synthetic model grid of Type~II SNe. The new model grid contains 228,016 Type~II SN models with different combinations of progenitor mass, explosion energy, \Ni\ mass, mass-loss rate, CSM radius, and CSM structure in much wider parameter ranges than the previous model grid of \citet{moriya2018}. We provide time evolution of spectral energy distributions (SEDs) from 1~\AA\ to 50,000~\AA\ at each time step for all the models. Thus, our model grid can be used to estimate light-curve evolution at any given photometric bands. Any extinction can also be directly applied to synthetic SEDs. In addition, our synthetic SEDs can be shifted at any redshifts so that the $K$ correction can be properly taken into account (see \cite{tominaga2011} for an example). We also provide photosphere information for all the models that is often required to break the degeneracy in estimating SN properties. A subset of this new model grid has already been successfully used to estimate the properties of 45 Type~II SNe discovered by ZTF \citep{subrayan2022}.

The rest of this paper is organized as follows. We present our model setups and numerical method used for our numerical calculations in Section~\ref{sec:setups}. We discuss some representative properties of our Type~II SN models in Section~\ref{sec:properties}. We provide a brief summary of this paper in Section~\ref{sec:summary}.

\begin{table}
  \tbl{Progenitor properties.}{%
  \begin{tabular}{cccc}
      \hline
      ZAMS Mass & Final Mass & H-rich envelope mass & Radius  \\ 
      \hline
      10~\Msun &  9.7~\Msun & 7.2~\Msun & 510~\Rsun \\
      12~\Msun & 10.9~\Msun & 7.8~\Msun & 640~\Rsun \\
      14~\Msun & 12.1~\Msun & 8.1~\Msun & 790~\Rsun \\
      16~\Msun & 13.1~\Msun & 8.4~\Msun & 890~\Rsun \\
      18~\Msun & 14.9~\Msun & 9.5~\Msun & 970~\Rsun \\
      \hline
    \end{tabular}}\label{tab:progenitors}
\end{table}

\begin{table*}
  \tbl{Parameters in the model grid.}{%
  \begin{tabular}{ll}
      \hline
      ZAMS Mass (\Msun) & $10,12,14,16,18$  \\
      $\dot{M}$ (\Msunpyr) & $10^{-5.0}, 10^{-4.5}, 10^{-4.0}, 10^{-3.5}, 10^{-3.0}, 10^{-2.5}, 10^{-2.0}, 10^{-1.5},10^{-1.0}$  \\
      $\beta$ & $0.5,1.0,1.5,2.0,2.5,3.0,3.5,4.0,4.5,5.0$ \\
      CSM radius ($10^{14}~\mathrm{cm}$) &  $1,2,4,6,8,10$ \\
      Explosion energy (B) & $0.5,1.0,1.5,2.0,2.5,3.0,3.5,4.0,4.5,5.0$ \\
      \Ni\ mass (\Msun) & $0.001,0.01,0.02,0.04,0.06,0.08,0.1,0.2,0.3$ \\
      \hline
    \end{tabular}}\label{tab:parameters}
\end{table*}

\section{Model setups}\label{sec:setups}
\subsection{Initial conditions}
\begin{figure}
  \begin{center}
	\includegraphics[width=\columnwidth]{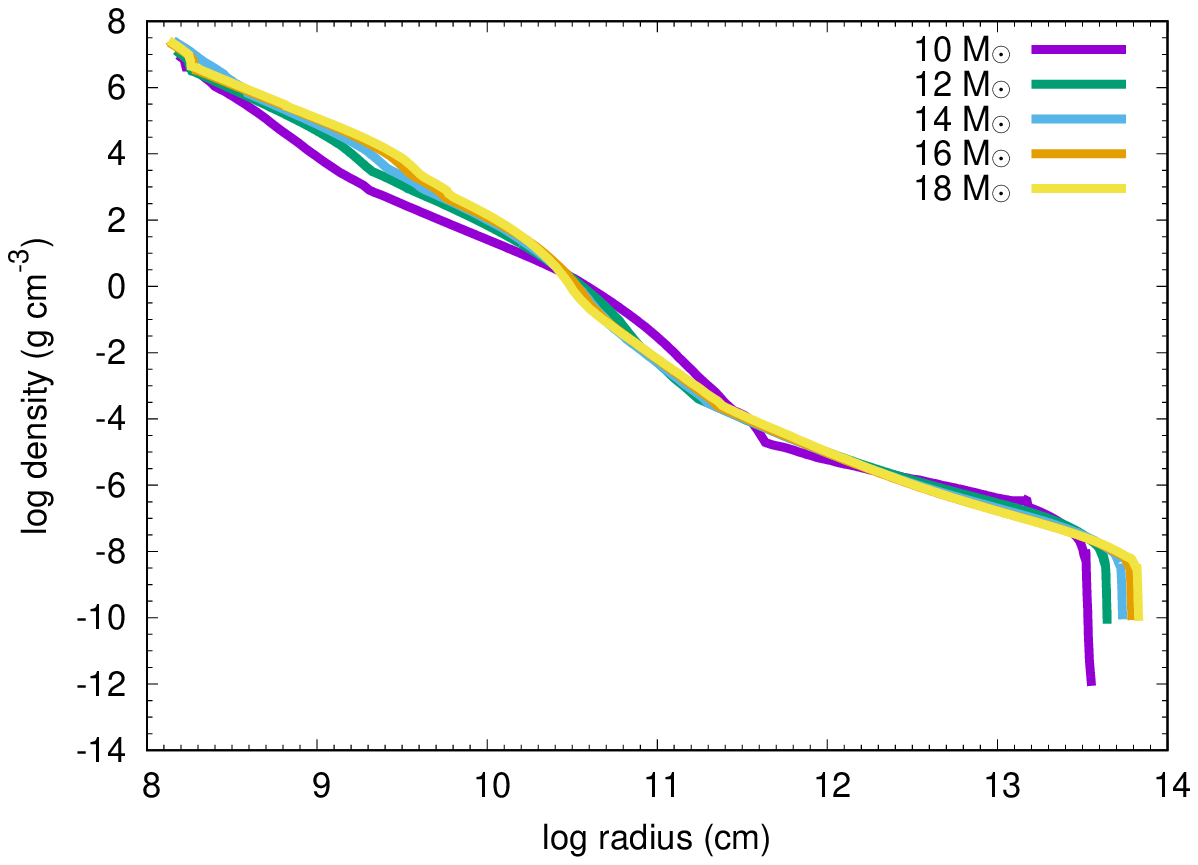}
	\includegraphics[width=\columnwidth]{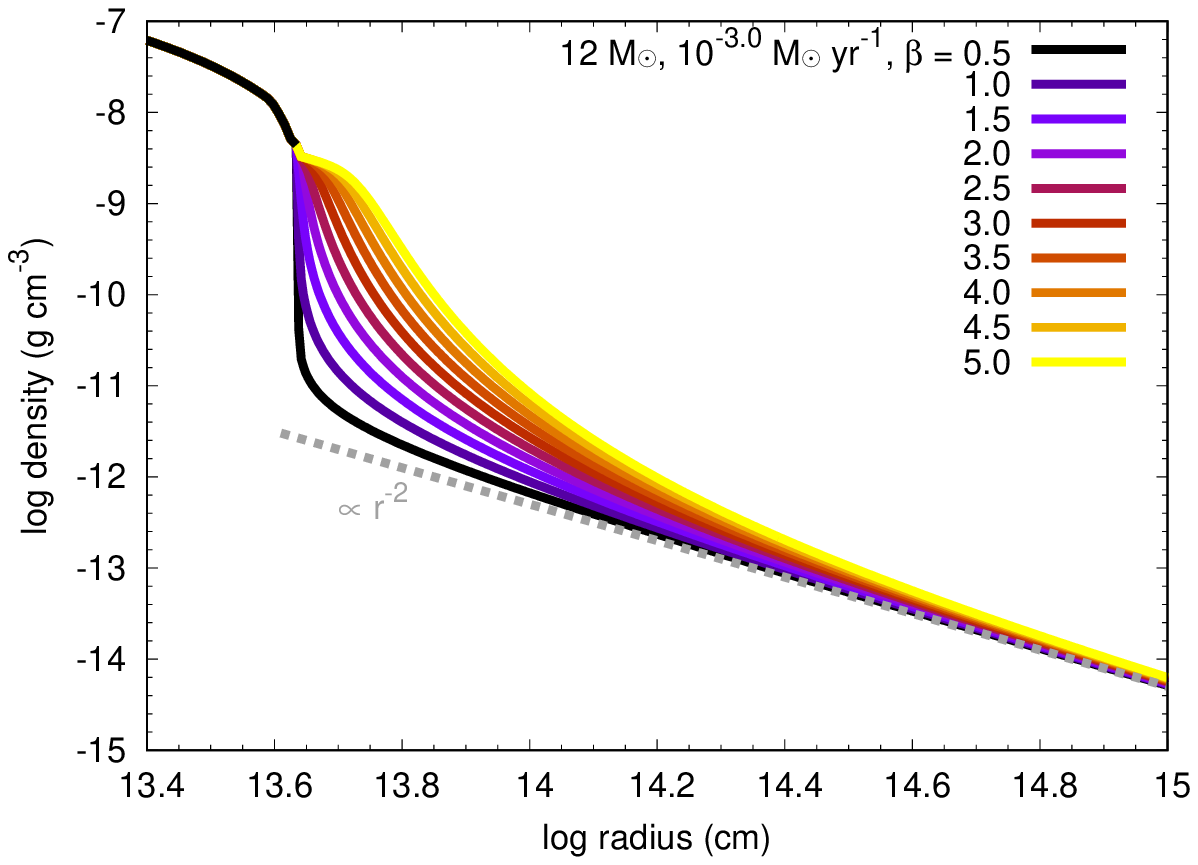}
  \end{center}
  \caption{%
  \textit{Top:} Density structure of the progenitor models above a mass cut of 1.4~\Msun. 
  \textit{Bottom:} Examples of CSM density structure with various $\beta$.
}%
  \label{fig:progenitor}
\end{figure}

We adopt red supergiant (RSG) Type~II SN progenitors in \citet{sukhbold2016}. We take progenitors with the zero-age main sequence (ZAMS) masses of $10,12,14,16,$ and $18~\Msun$, which is in the estimated mass range of Type~II SN progenitors (e.g., \cite{smartt2015,davies2018,davies2020,kochanek2020,rodriguez2022}). They have solar metallicity at ZAMS. The progenitor properties are summarized in Table~\ref{tab:progenitors}. We assume a mass cut of 1.4~\Msun, and put the structure above the mass cut in our explosion calculations. The density structure of the progenitor models is presented in Fig.~\ref{fig:progenitor}. 

We attach confined CSM on top of the progenitors. It has been recently recognized that confined dense CSM often exists around Type~II SN progenitors at the time of explosion (e.g., \cite{yaron2017,bruch2021,bruch2022,jacobson-galan2022}), and it strongly affect the early phases of Type~II SN properties (e.g., \cite{moriya2011,morozova2018,haynie2021,kozyreva2022,dessart2022}). The origin and structure of the confined CSM are still unknown (e.g., \cite{yoon2010,fuller2017,dessart2017,morozova2020,tsuna2021,ko2022}). In this work, we adopt the following CSM density structure based on the previous studies by \citet{moriya2017,moriya2018},
\begin{equation}
\rho_\mathrm{CSM}(r) = \frac{\dot{M}}{4\pi v_\mathrm{wind}(r)}r^{-2},
\end{equation}
where $r$ is the radius, $\dot{M}$ is the mass-loss rate, and $v_\mathrm{wind}$ is the wind velocity. For the wind velocity, we adopt 
\begin{equation}
    v_\mathrm{wind}(r) = v_0 + \left(v_\infty-v_0\right)\left(1-\frac{R_0}{r}\right)^{\beta},
\end{equation}
where $v_0$ is an initial wind velocity at a progenitor surface, $v_\infty$ is a terminal wind velocity, $R_0$ is a progenitor radius, and $\beta$ is a wind structure parameter determined by the efficiency of wind acceleration. RSGs are known to have slow wind acceleration with $\beta> 1$ (e.g., \cite{schroeder1985,baade1996}). This expression assumes that confined CSM around Type~II SN progenitors is caused by strong stellar wind. Although it is still not clear if such a strong wind model is responsible for the formation of confined CSM around Type~II SNe \citep{davies2022}, it provides a good explanation for early photometric properties of Type~II SNe \citep{forster2018}. We choose $v_0$ so that CSM structure is smoothly connected at the progenitor surface. $v_\infty$ is set to 10~\kmps, which is a typical wind velocity of RSGs (e.g., \cite{mauron2011,goldman2017}). In our model grid, we change $\dot{M}$ and $\beta$. We adopt $\dot{M}=10^{-5.0}, 10^{-4.5}, 10^{-4.0}, 10^{-3.5}, 10^{-3.0}, 10^{-2.5}, 10^{-2.0}, 10^{-1.5},$ and $10^{-1.0}~\Msunpyr$ and $\beta=0.5,1.0,1.5,2.0,2.5,3.0,3.5,4.0,4.5,$ and $5.0$. The radius of the confined CSM is constrained to be less than around $10^{15}~\mathrm{cm}$ \citep{yaron2017}. Thus, we adopt the CSM radius of $10^{14}, 2\times 10^{14}, 4\times 10^{14}, 6\times 10^{14}, 8\times 10^{14},$ and $10^{15}~\mathrm{cm}$. Figure~\ref{fig:progenitor} presents some examples of CSM density structure with various $\beta$. The CSM abundance is set to be the same as that at the progenitor surface.

\subsection{Explosion properties}
We adopt explosion energies of $0.5,1.0,1.5,2.0,2.5,3.0,3.5,4.0,4.5,$ and 5.0~B where $1~\mathrm{B}\equiv 10^{51}~\mathrm{erg}$. While explosion energy of Type~II SNe is typically estimated to be $0.1-1.5~\mathrm{B}$ (e.g., \cite{martinez2022iii}), we extend our models to higher explosion energy because magnitude limited transient surveys tend to discover SNe with large luminosity having high explosion energy.

The mass of \Ni\ strongly affects light-curve properties of Type~II SNe at late phases. In our model grid, we adopt \Ni\ mass of $0.001,0.01,0.02,0.04,0.06,0.08,0.1,0.2,$ and $0.3~\Msun$, which is in the range of \Ni\ mass estimated from observations (e.g., \cite{anderson2019}). One important property that is related to \Ni\ and determines Type~II light-curve features is the degree of \Ni\ mixing (e.g., \cite{kasen2009,bersten2011,dessart2013,moriya2016}). Although it is likely that \Ni\ is mixed to hydrogen-rich envelopes in Type~II SNe (e.g., \cite{singh2019}), it is not exactly clear how far in the hydrogen-rich envelopes \Ni\ is mixed. In our model grid, we assume that \Ni\ is uniformly mixed up to the half mass of the hydrogen-rich envelope so that an average effect of \Ni\ mixing is taken into account.

We have introduced all the parameters in our model grid that is summarized in Table~\ref{tab:parameters}. If we take all the parameter combinations, we should have 243,000 models in total in our model grid. However, computations of some models failed for various numerical reasons. Failed models are not concentrated on particular parameter ranges. We succeeded in completing the numerical calculations for 228,016 models which are 94\% of the total parameter combination. We introduce our numerical methods in the next section.

% s10 46391
% s12 44974
% s14 44995
% s16 46417
% s18 45239
% total 228016 (0.938337)
% 48600 x 5 = 243000

\subsection{Numerical computations}
We adopt the one-dimensional multi-frequency radiation hydrodynamics code \texttt{STELLA} to construct the model grid \citep{blinnikov1998,blinnikov2000,blinnikov2006}. \texttt{STELLA} has been used for Type~II SN light-curve modeling intensively (e.g., \cite{goldberg2019,hiramatsu2021,kozyreva2022b} for recent examples). \texttt{STELLA} provides synthetic light curves that are consistent with other radiation transport codes used in SN modeling \citep{blondin2022}. Being a radiation hydrodynamics code, \texttt{STELLA} can take the interaction between SN ejecta and CSM into account in computing light curves. We introduce \texttt{STELLA} briefly.

\texttt{STELLA} evaluates the evolution of hydrodynamic and radiation parameters in each time step by using the variable Eddington factor method. Explosions are initiated by injecting thermal energy required to achieve each explosion energy above the mass cut at 1.4~\Msun\ in the first $0.1~\mathrm{sec}$. Radiation is treated in multi-frequency, and \texttt{STELLA} provides a SED at each time step. We adopt the standard 100 frequency bins from 1~\AA\ to 50,000~\AA\ in a log scale. Opacity in each frequency bin is estimated based on the atomic levels from the Saha equation by assuming the local thermodynamic equilibrium. Bremsstrahlung, bound-free transitions, bound-bound transitions, photoionization, and electron scattering are taken into account in opacity. Gas temperature and radiation temperature are treated separately. We define photosphere as a location where a Rosseland-mean optical depth becomes $2/3$. We record radius, velocity, mass coordinate, and temperature at the photosphere. 

Bolometric light curves are computed by integrating SEDs at each time step. Light curves in any photometric bands can be obtained by convolving SEDs with filter response functions. We provide light curves for the \textit{u}, \textit{g}, \textit{r}, \textit{i}, \textit{z}, and \textit{y} filters of LSST\footnote{\url{https://github.com/lsst/throughputs/tree/main/baseline}}, the \textit{g}, \textit{r}, and \textit{i} filters of ZTF \citep{bellm2019}, the \textit{g}, \textit{r}, and \textit{i} filters of SDSS \citep{doi2010}, and the \textit{UVW2}, \textit{UVM2}, \textit{UVW1}, \textit{U}, \textit{B}, and \textit{V} filters of the Neil Gehrels Swift Observatory \citep{poole2008,breeveld2011} at the rest frame in the absolute AB magnitude. The absolute AB magnitude at a \textit{x} filter is evaluated by
\begin{equation}
    M_{\mathrm{AB},x} = -2.5\log\frac{\int F_\nu(\nu)S_x(\nu)d\nu/\nu}{\int S_x(\nu)d\nu/\nu} -48.60,
\end{equation}
where $\nu$ is the frequency, $F_\nu(\nu)$ is the absolute synthetic flux observed at 10~pc in $\mathrm{erg~s^{-1}~cm^{-2}~Hz^{-1}}$, and $S_x(\nu)$ is a filter response function (e.g., \cite{bessell2012}).

\section{Some light-curve properties}\label{sec:properties}
The model grid computed by the method discussed so far is available at \url{https://doi.org/10.5061/dryad.pnvx0k6sj}. We present some features of Type~II SNe predicted by the model grid in this section.

\begin{figure}
  \begin{center}
	\includegraphics[width=\columnwidth]{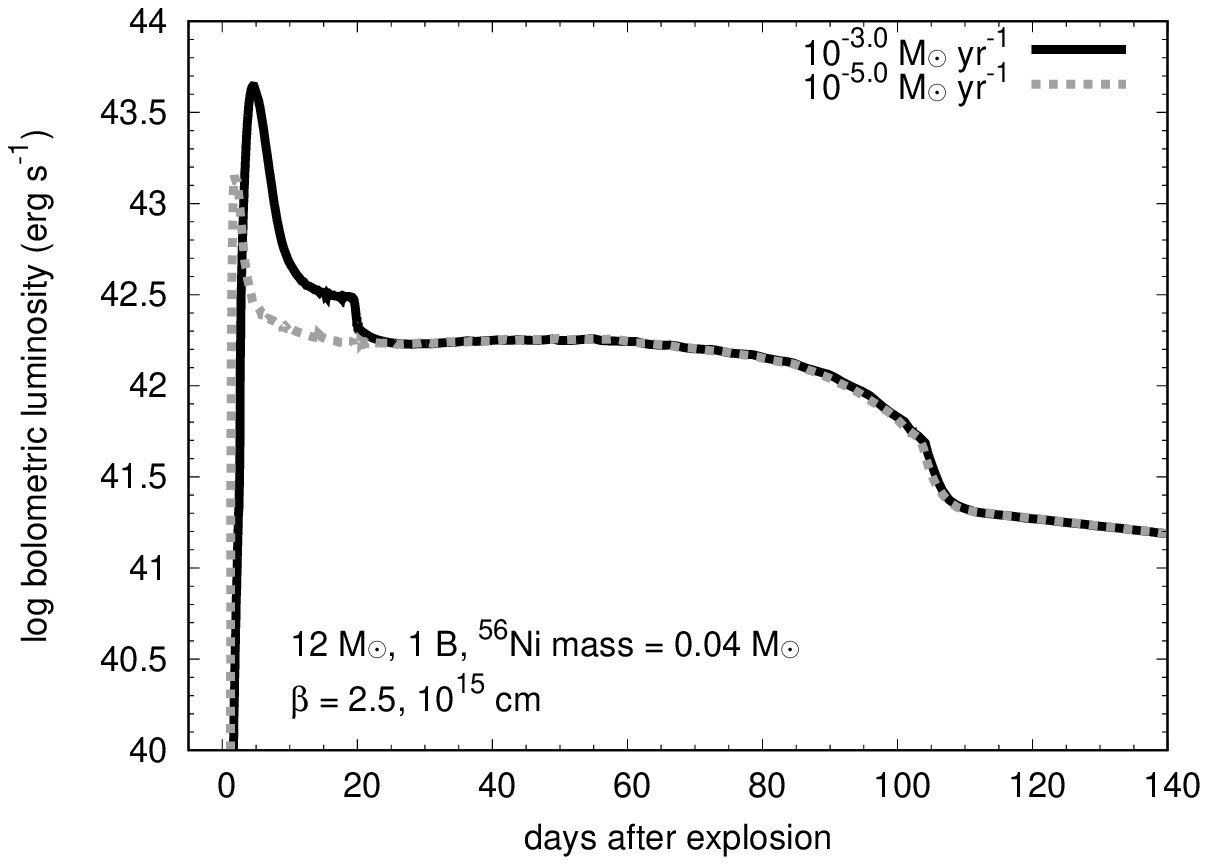}
	\includegraphics[width=\columnwidth]{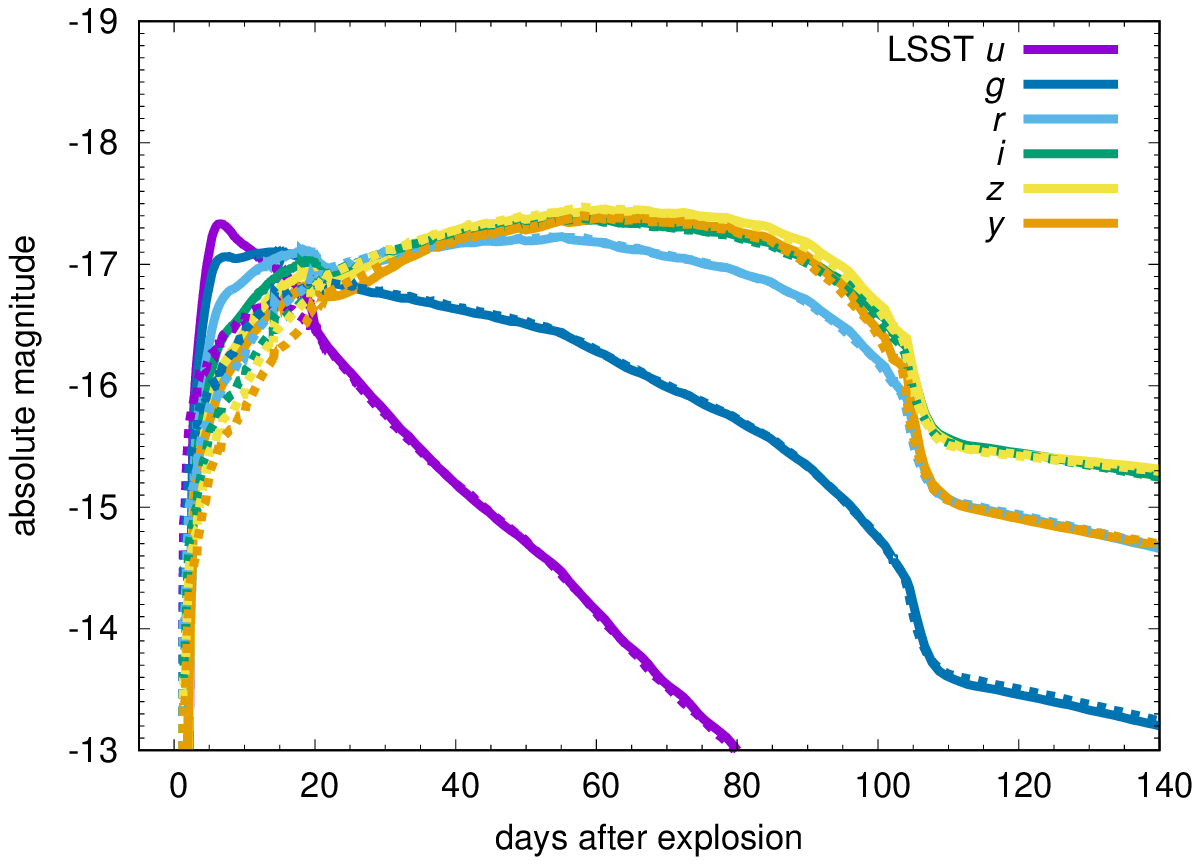}
	\includegraphics[width=\columnwidth]{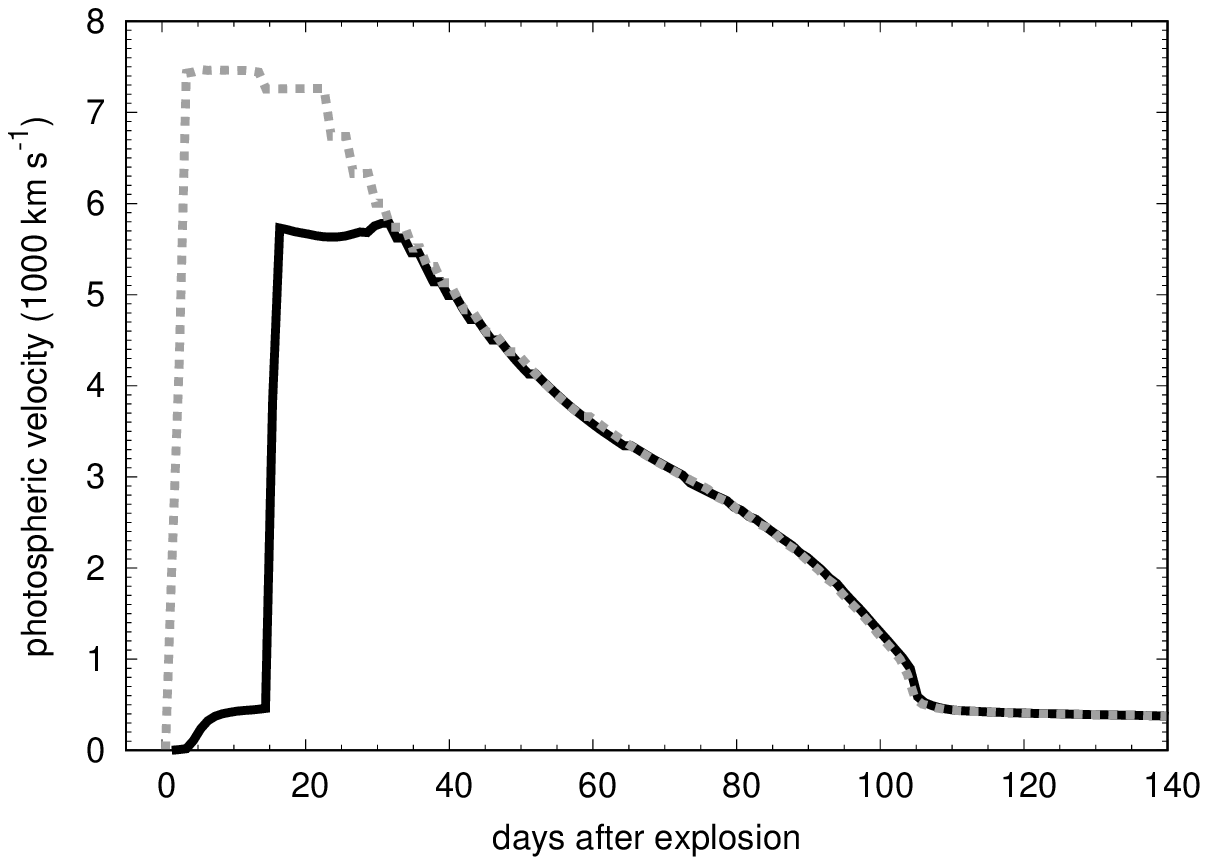}
  \end{center}
  \caption{%
  Examples of light curves and photospheric velocities with low ($10^{-5.0}~\Msunpyr$, dashed lines) and high ($10^{-3.0}~\Msunpyr$, solid lines) mass-loss rates in the model grid. The two models have the same progenitor ($12~\Msun$), explosion energy ($1~\mathrm{B}$), \Ni\ mass ($0.04~\Msun$), $\beta$ ($2.5$), and CSM radius ($10^{15}~\mathrm{cm}$). The top panel presents bolometric light curves, the middle panel shows optical light curves in the LSST filters, and the bottom panel shows photospheric velocities.
}%
  \label{fig:examples}
\end{figure}

Figure~\ref{fig:examples} shows examples of our model grid data. Light curves and photospheric velocity from two different mass-loss rates ($10^{-5.0}$ and $10^{-3.0}~\Msunpyr$) are presented. We can see clearly the effects of dense confined CSM in early phases in these examples. The early bolometric luminosity becomes higher when the mass-loss rate is higher because of extra radiation from CSM interaction. The rise time in bolometric light curves increases with higher mass-loss rates due to increased diffusion time in CSM. The rise times in the optical bands become shorter with higher mass-loss rates because contributions of the luminosity from CSM interaction becomes dominant in optical in early phases. We can find that photospheric velocity is also strongly affected by the CSM interaction in early phases while the CSM interaction continues. The photosphere can be located within CSM in early phases and the photospheric velocity can be low at the beginning. We can observe the effect of acceleration of CSM by radiation in the photospheric velocity before the shock passes through the CSM (cf. \cite{tsuna2023}).

\begin{figure*}
  \begin{center}
	\includegraphics[width=\columnwidth]{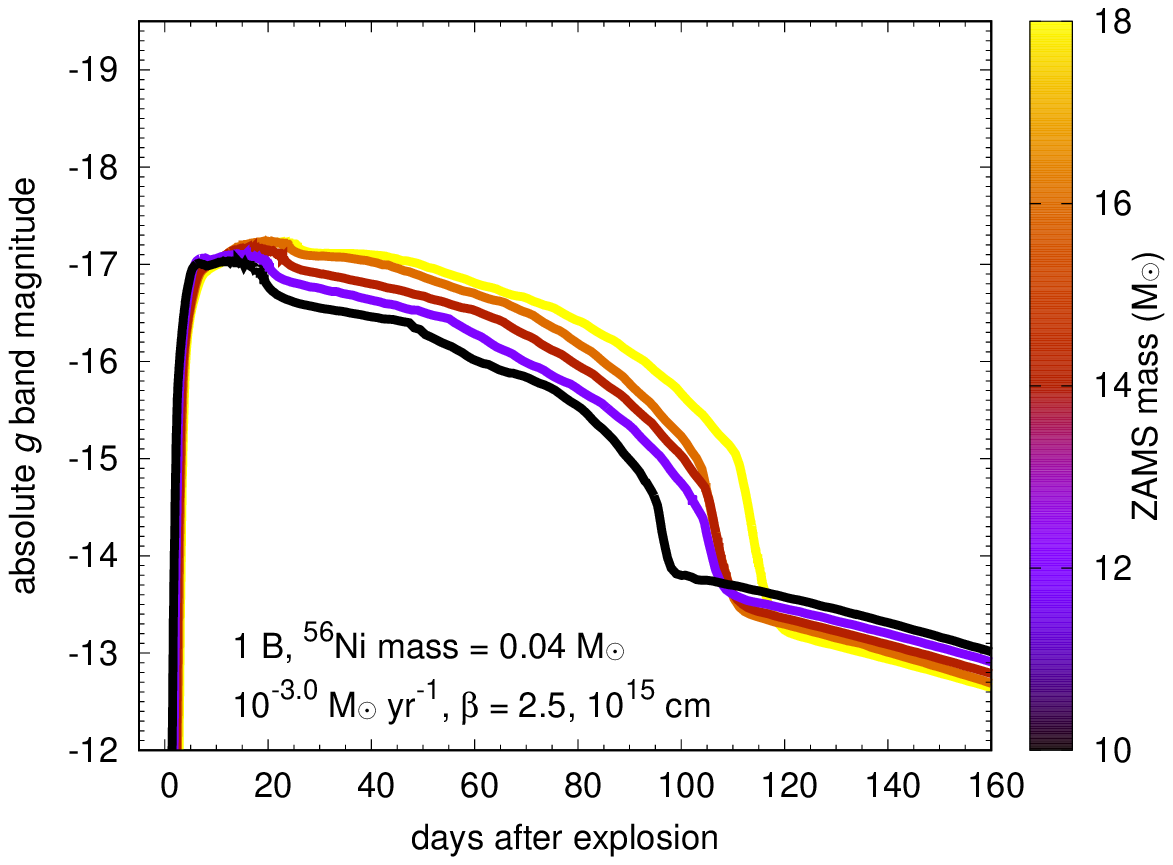}
	\includegraphics[width=\columnwidth]{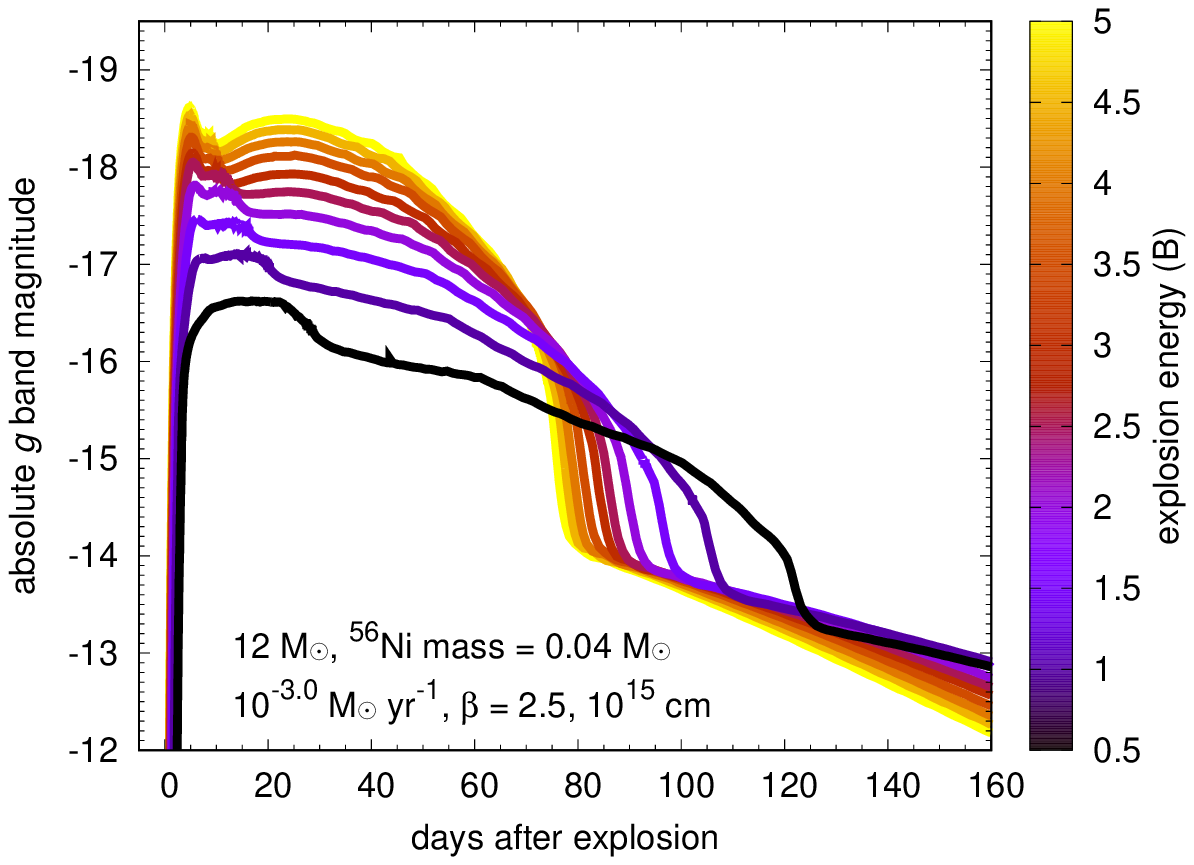}
	\includegraphics[width=\columnwidth]{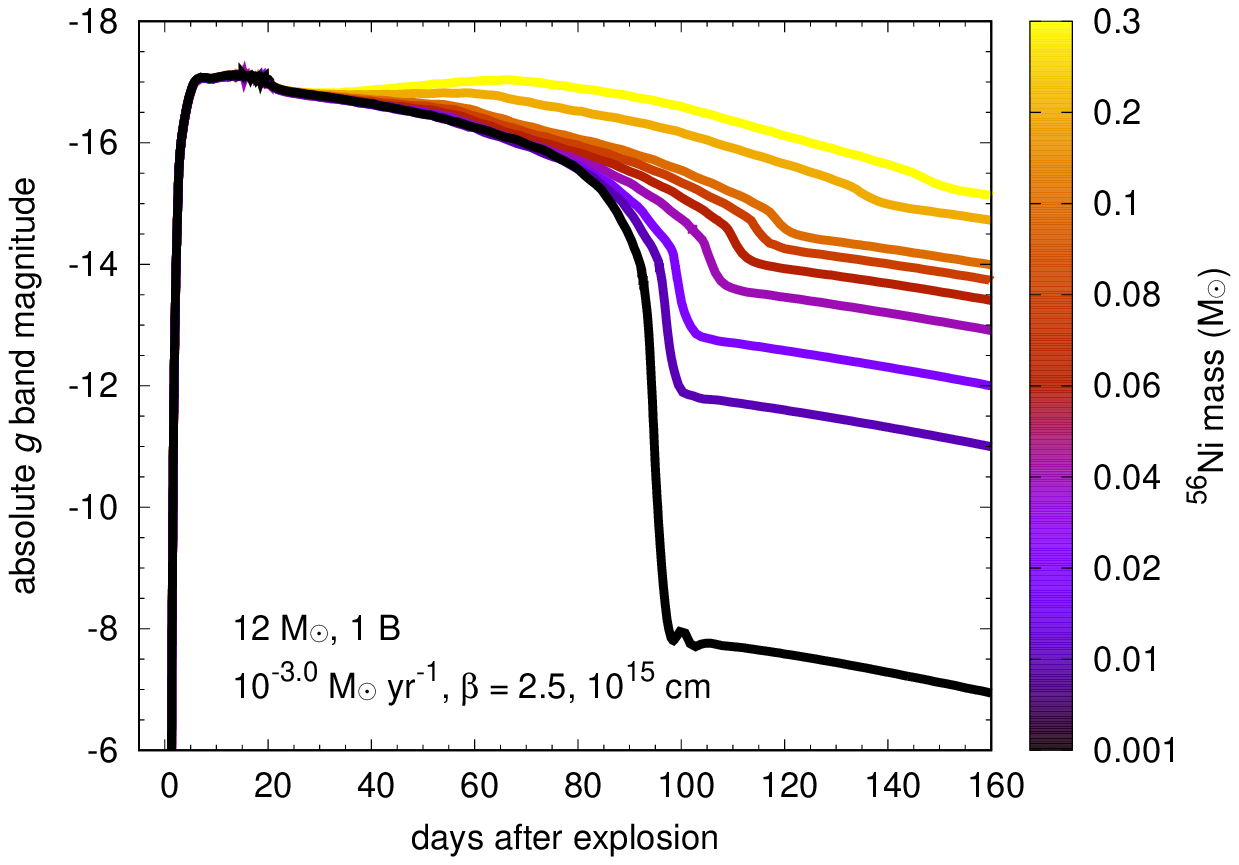}
	\includegraphics[width=\columnwidth]{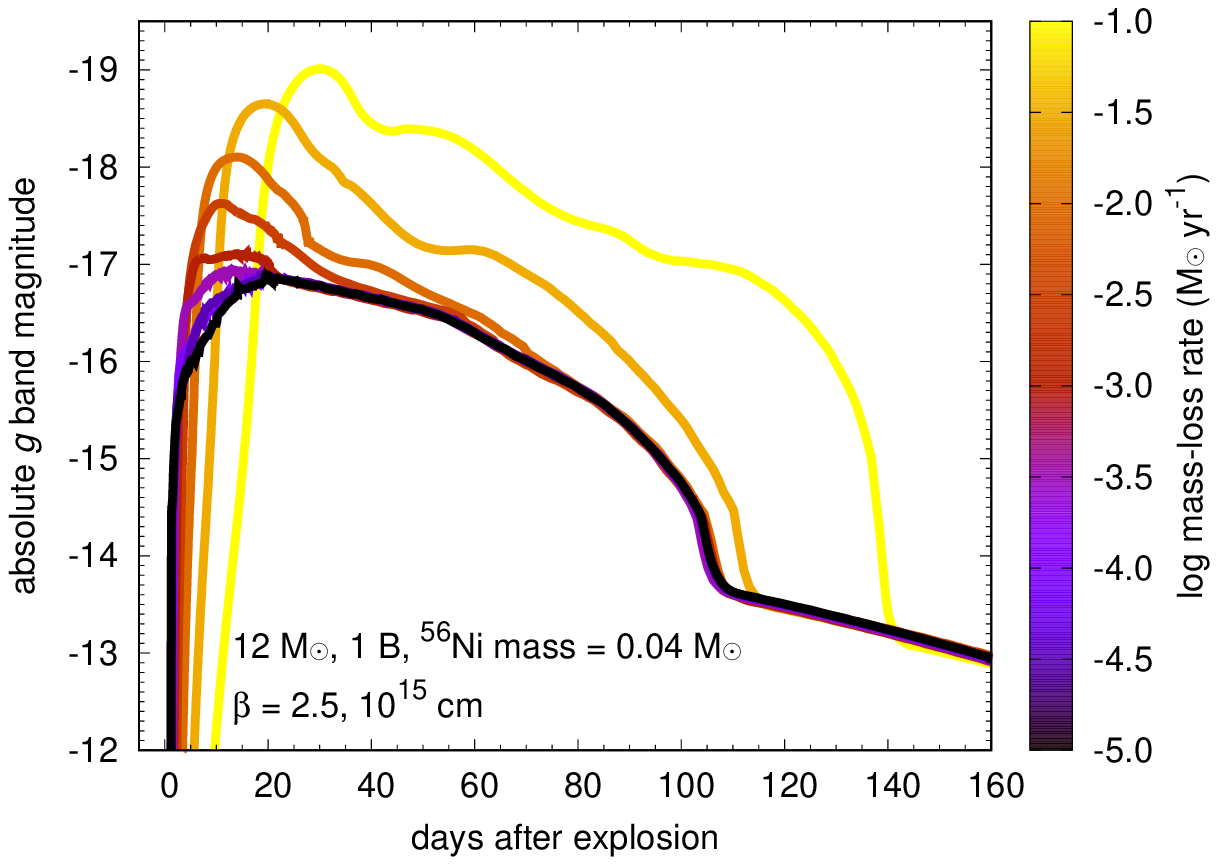}
	\includegraphics[width=\columnwidth]{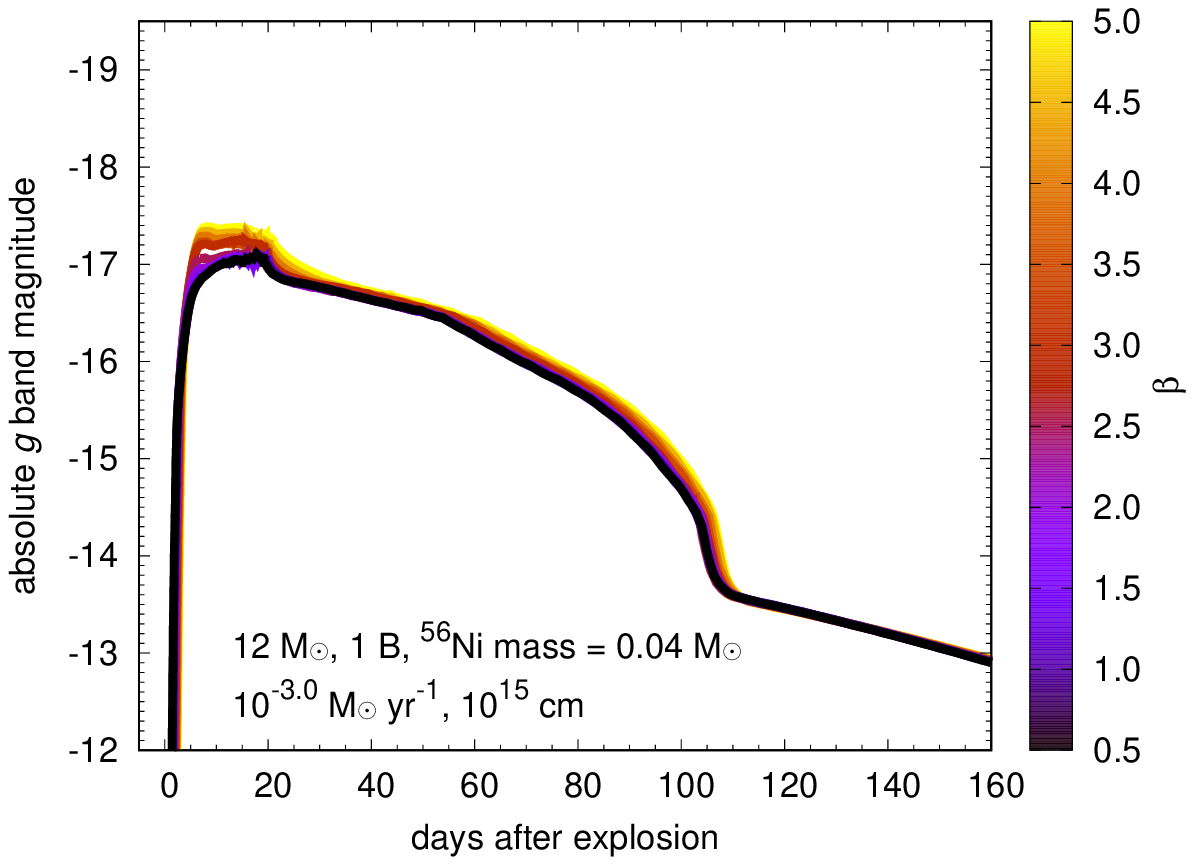}
	\includegraphics[width=\columnwidth]{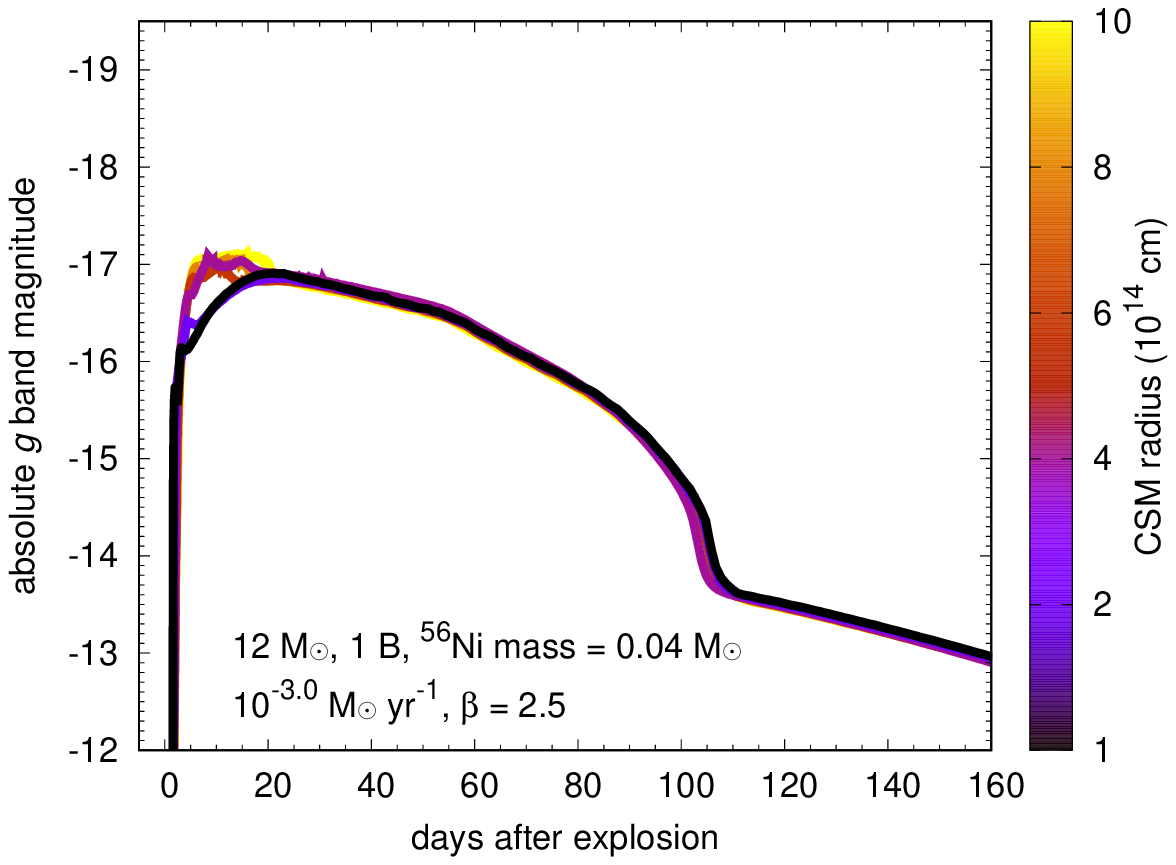}
  \end{center}
  \caption{%
  Effects of each parameter on the LSST \textit{g} band light curves. Each panel shows the effects of one parameter by only changing it. The other parameters are fixed in each panel.
}%
  \label{fig:gband}
\end{figure*}

Figure~\ref{fig:gband} illustrates how each parameter affects light-curve properties in the LSST $g$ band. The rise times of Type~II SNe are mostly affected by CSM properties and explosion energy. The tail phase is mostly affected by \Ni\ mass, although explosion energy can also change the light-curve decline rate in the tail phase. The plateau magnitude and duration are affected by progenitor mass, explosion energy, and \Ni\ mass. Mass-loss rates also affect the plateau properties when they are high enough.

\begin{figure*}
  \begin{center}
	\includegraphics[width=0.66\columnwidth]{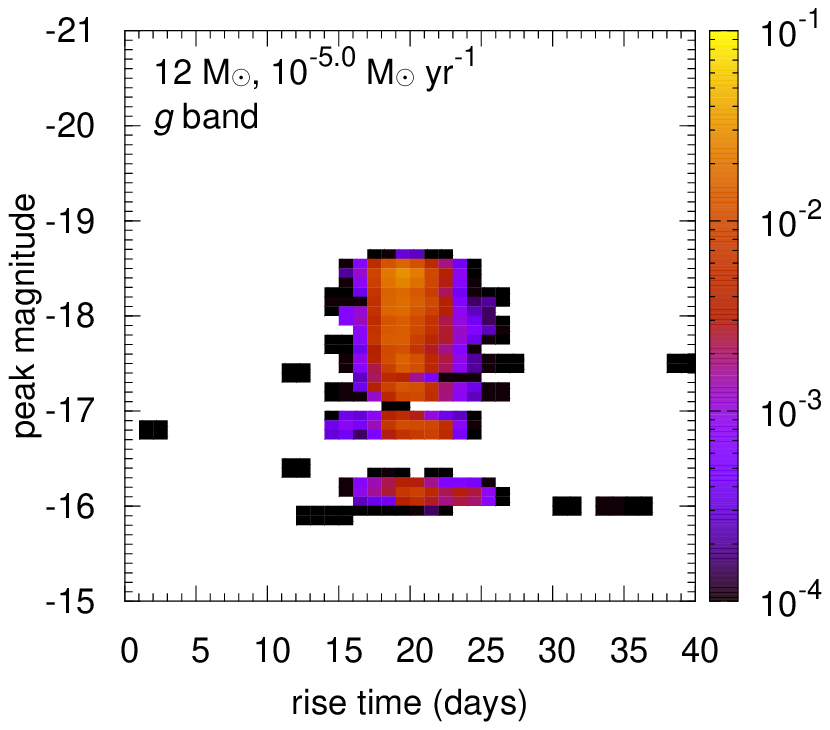}
	\includegraphics[width=0.66\columnwidth]{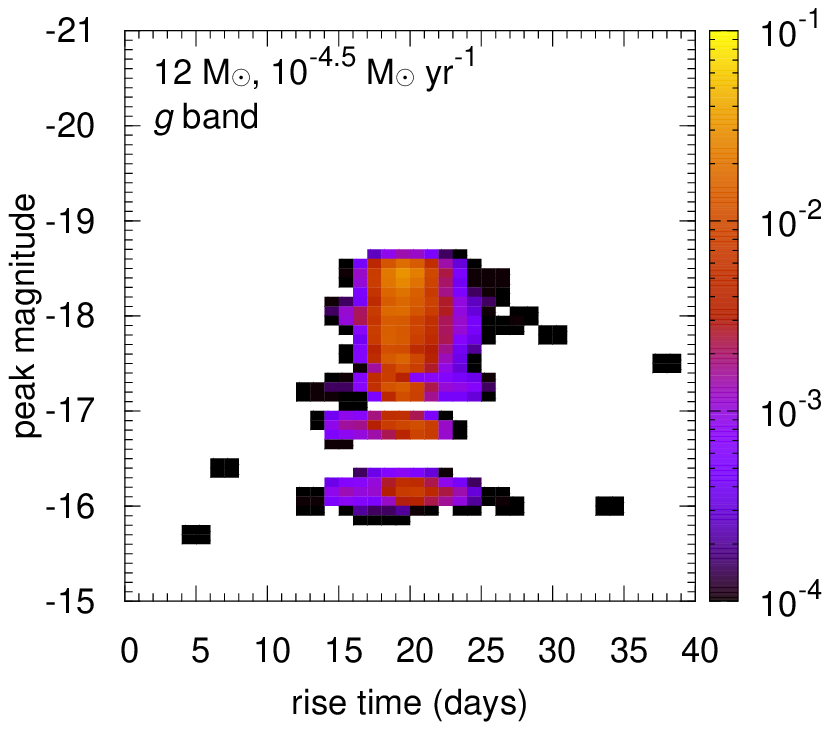}
	\includegraphics[width=0.66\columnwidth]{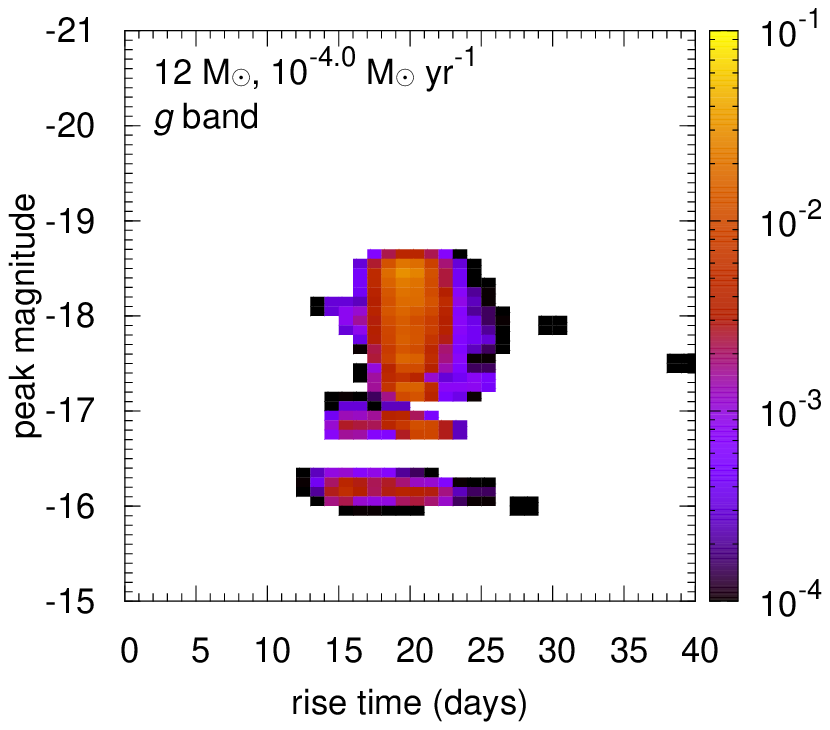}
	\includegraphics[width=0.66\columnwidth]{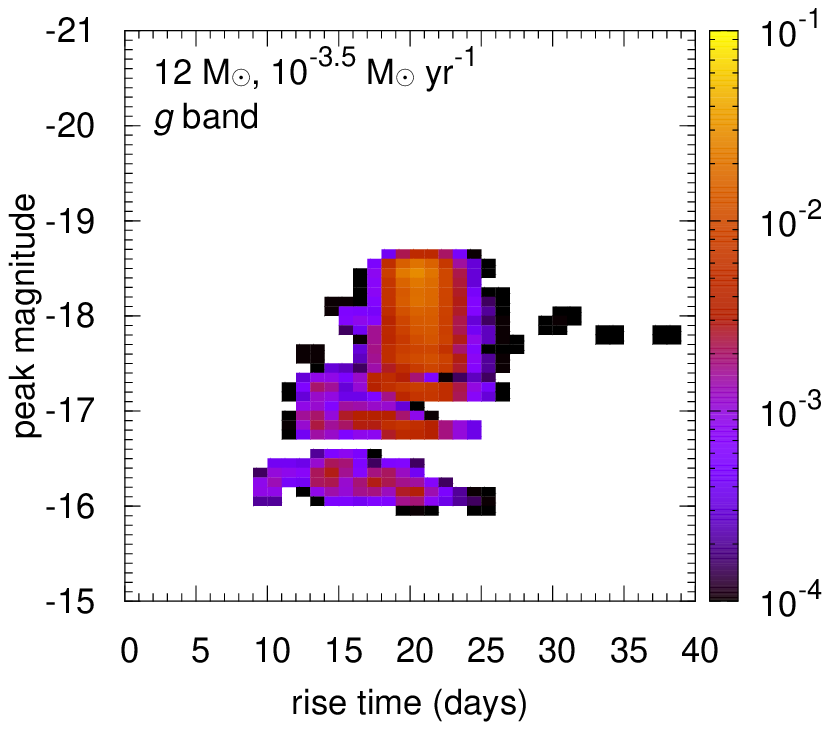}
	\includegraphics[width=0.66\columnwidth]{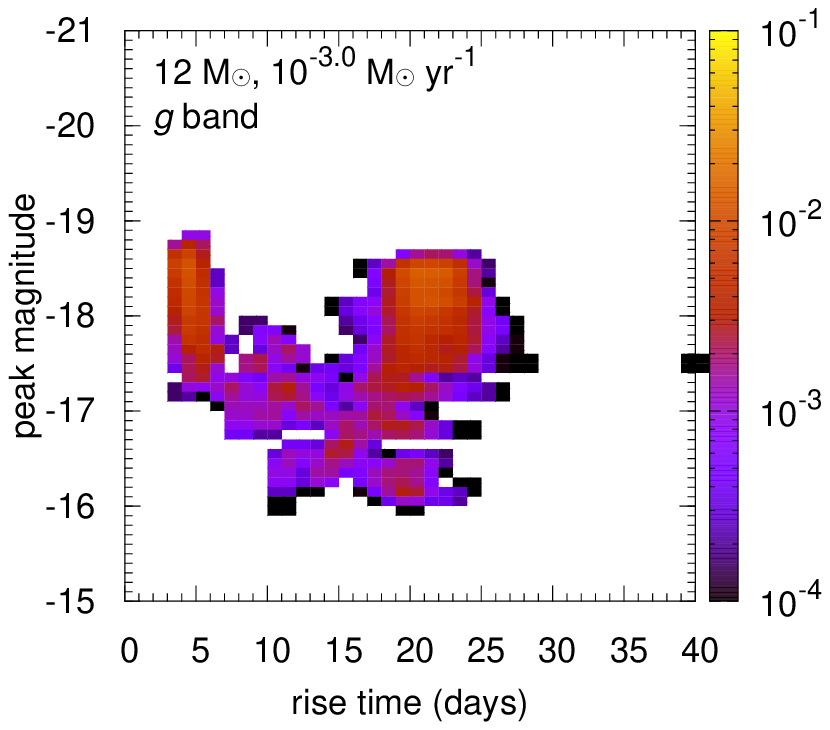}
	\includegraphics[width=0.66\columnwidth]{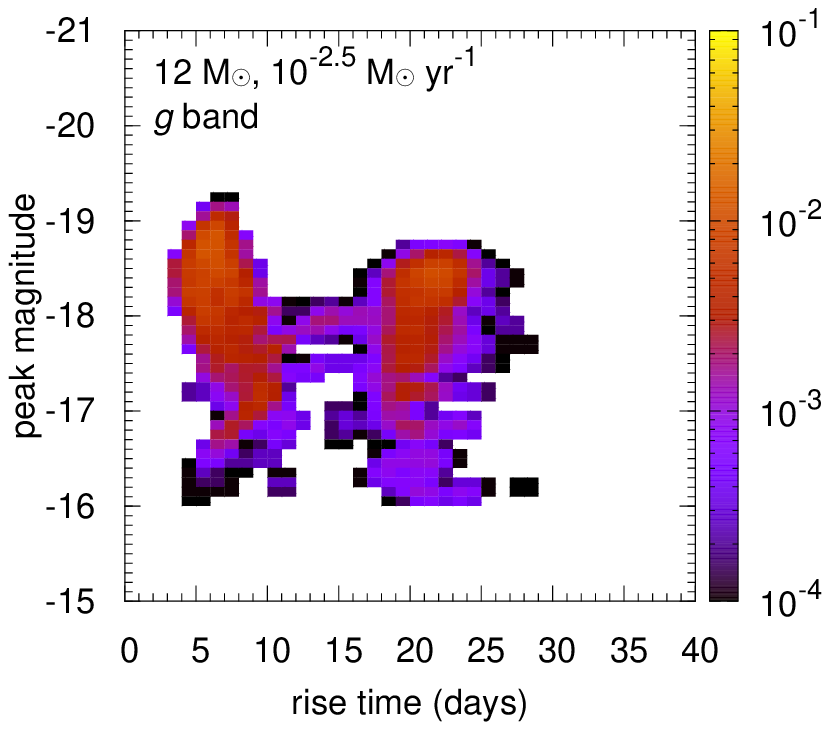}
	\includegraphics[width=0.66\columnwidth]{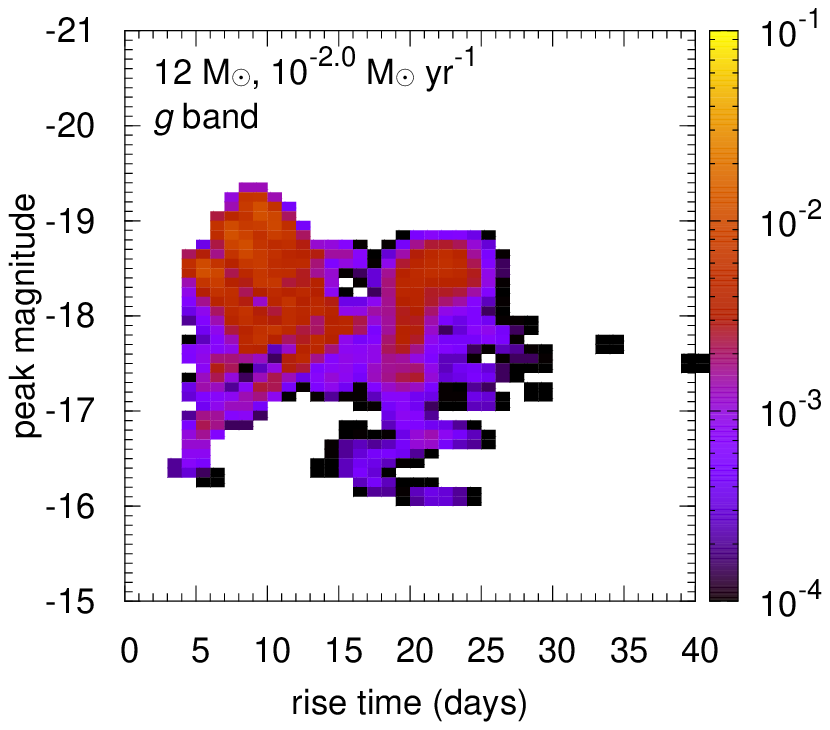}
	\includegraphics[width=0.66\columnwidth]{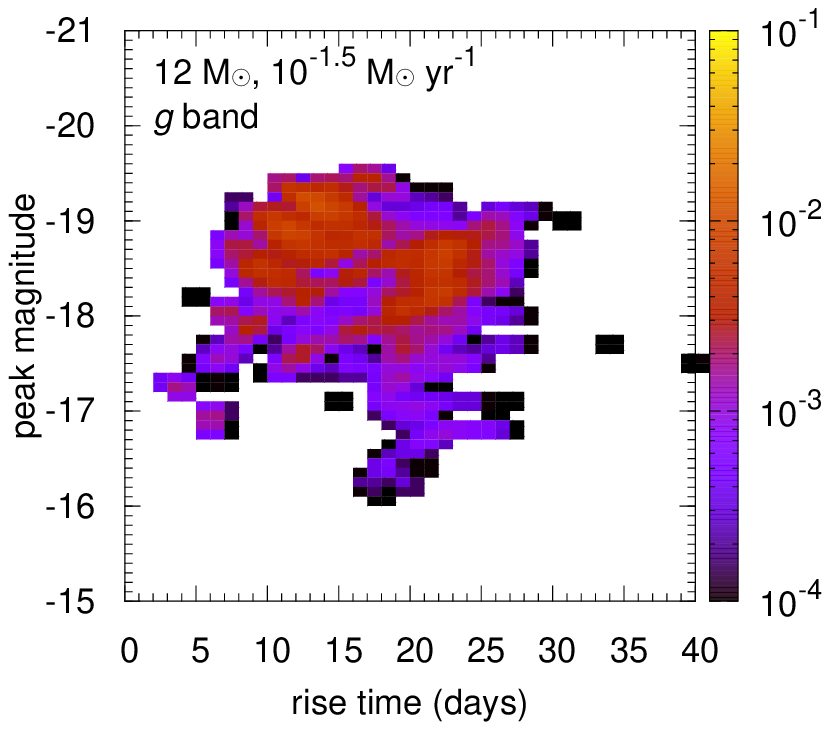}
	\includegraphics[width=0.66\columnwidth]{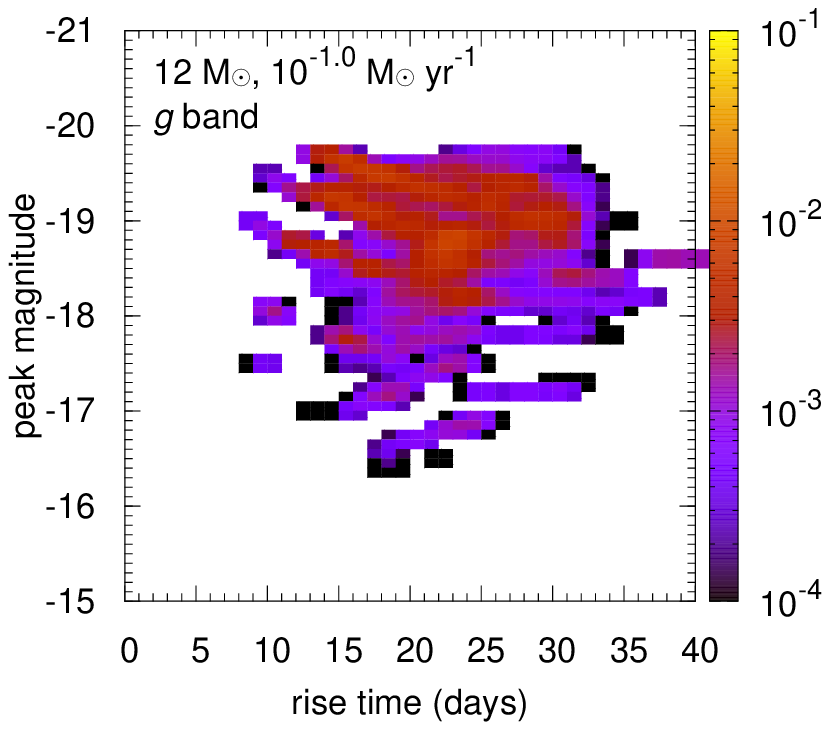}
  \end{center}
  \caption{%
  Rise time and peak luminosity distributions of Type~II SNe in the LSST \textit{g} band from the 12~\Msun\ progenitor. The color contours show relative fractions of the models within each bin. Each panel shows a summary of all the models with one mass-loss rate. 
}%
  \label{fig:s12_risepeak_g_mdot}
\end{figure*}

\begin{figure*}
  \begin{center}
	\includegraphics[width=0.66\columnwidth]{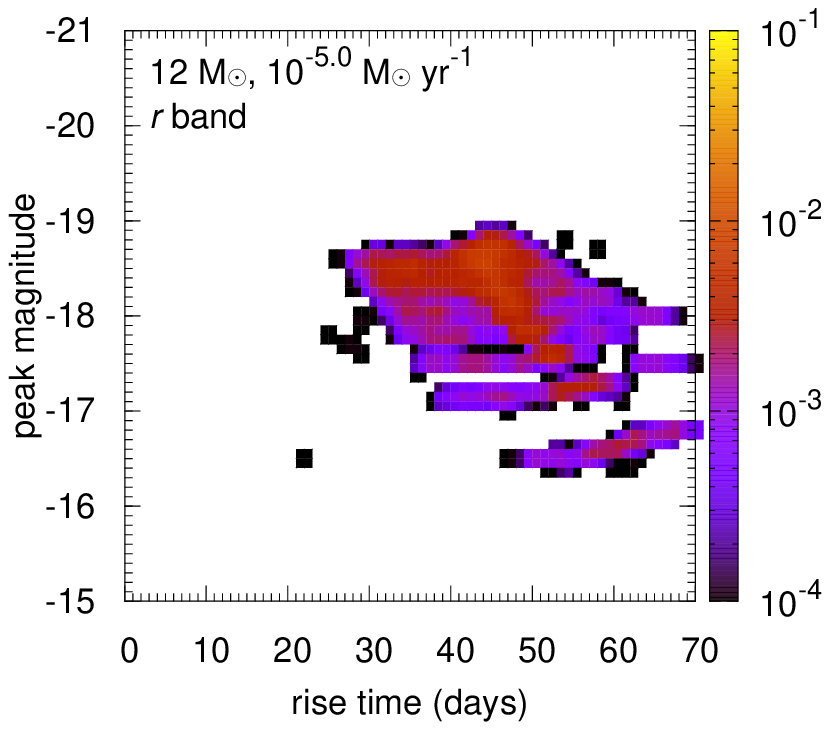}
	\includegraphics[width=0.66\columnwidth]{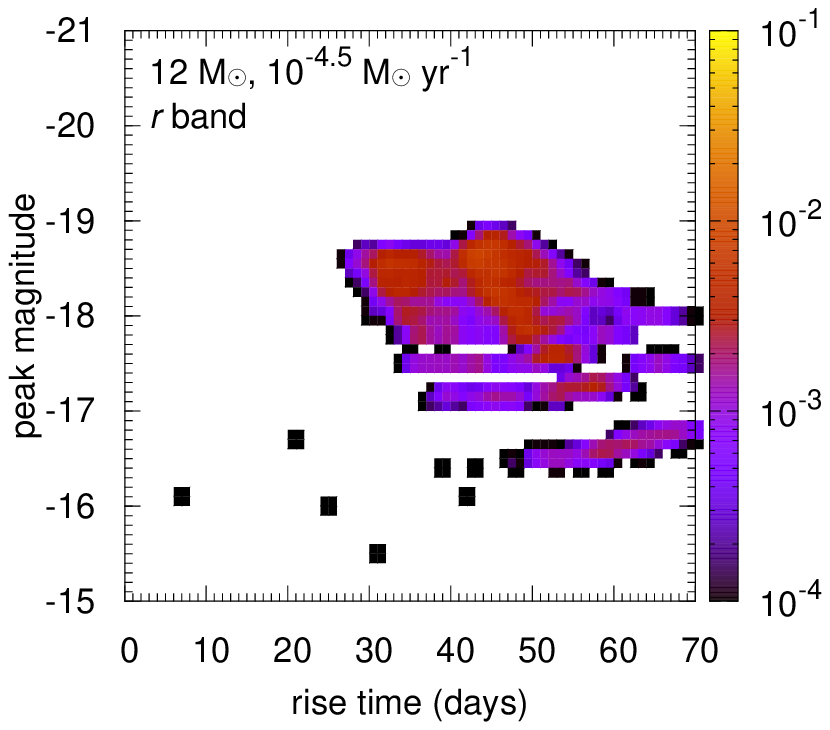}
	\includegraphics[width=0.66\columnwidth]{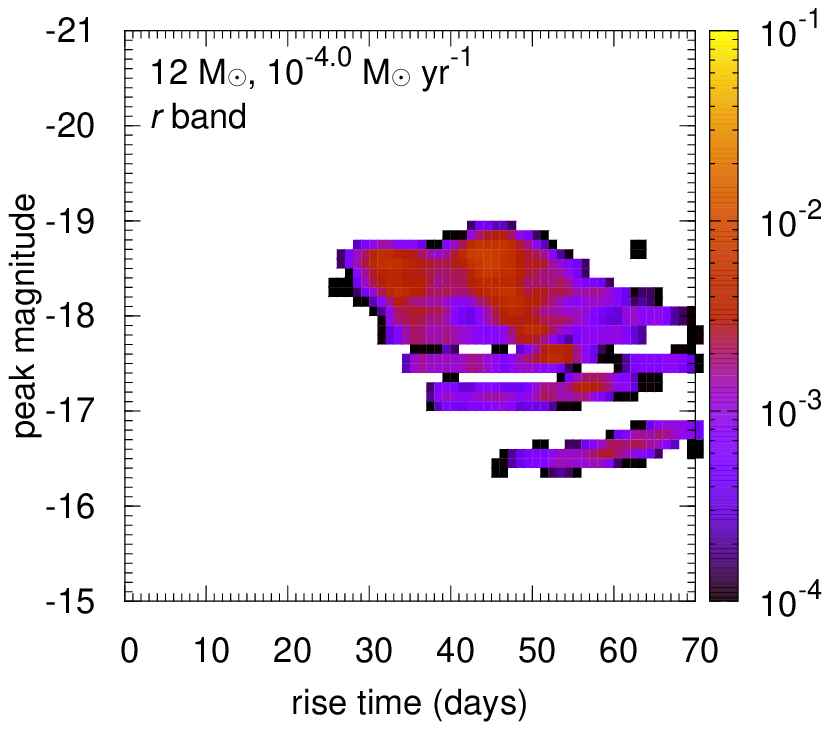}
	\includegraphics[width=0.66\columnwidth]{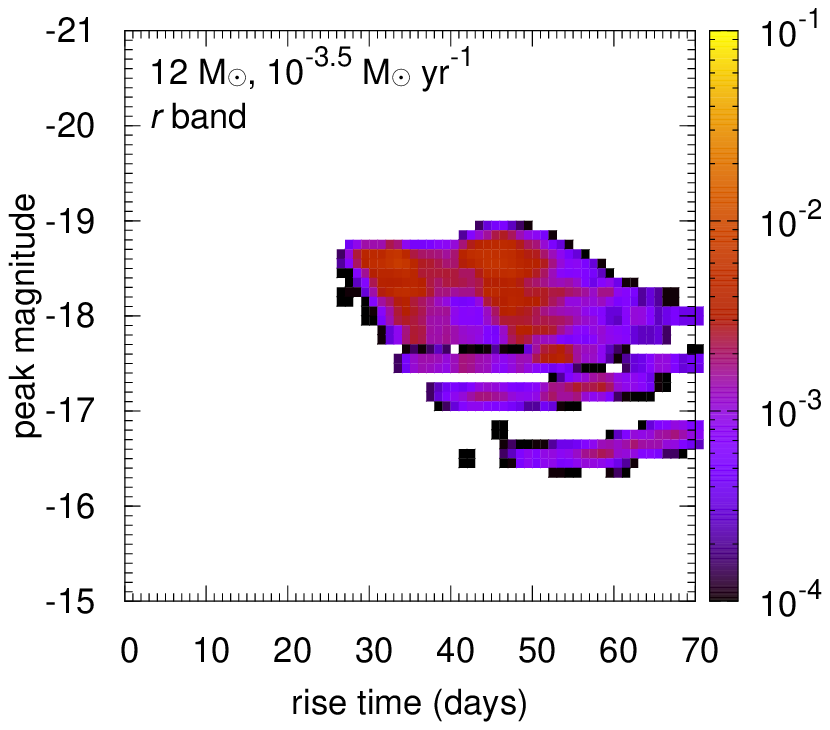}
	\includegraphics[width=0.66\columnwidth]{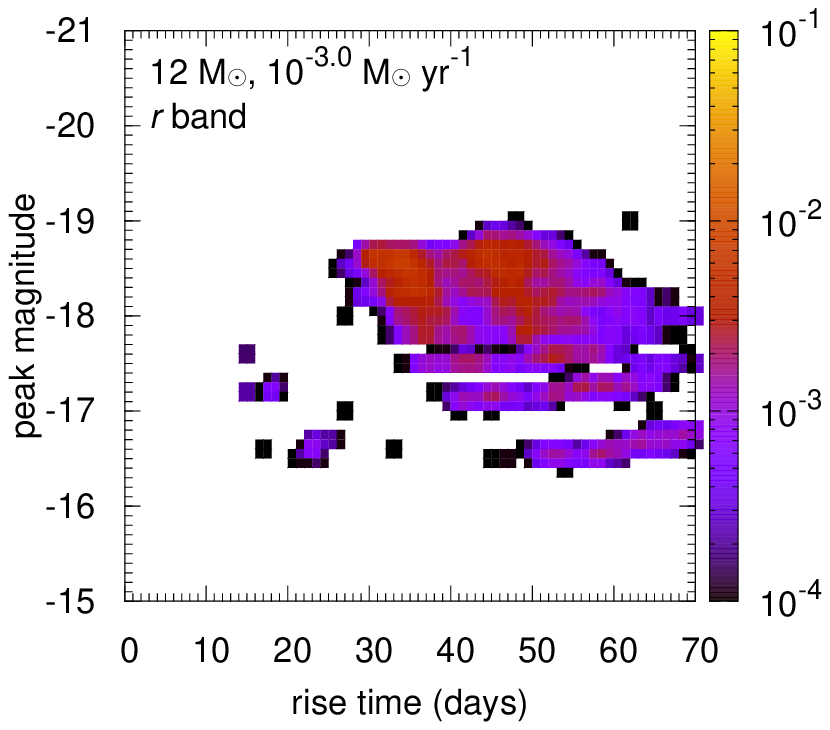}
	\includegraphics[width=0.66\columnwidth]{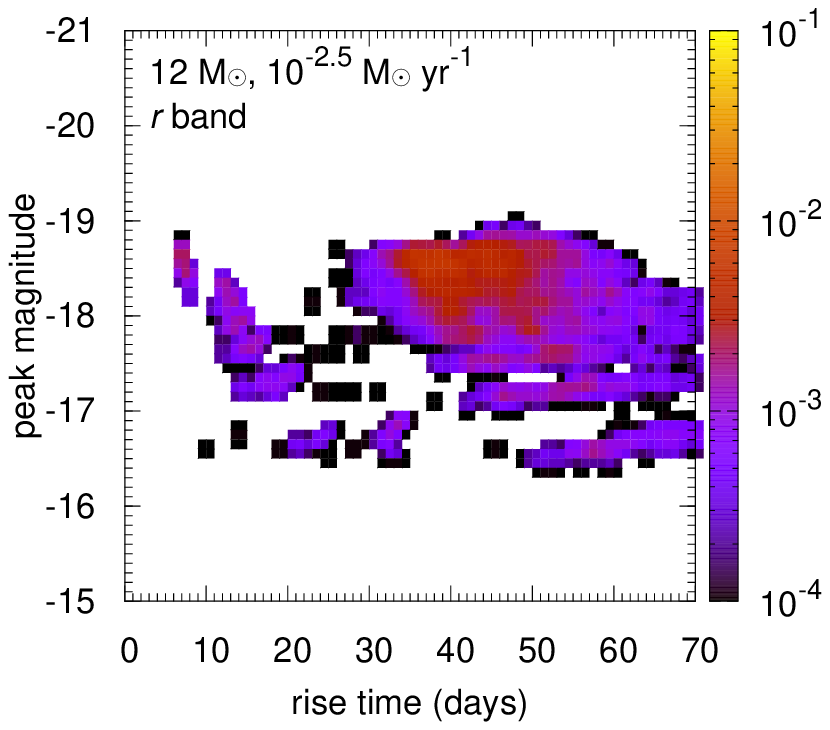}
	\includegraphics[width=0.66\columnwidth]{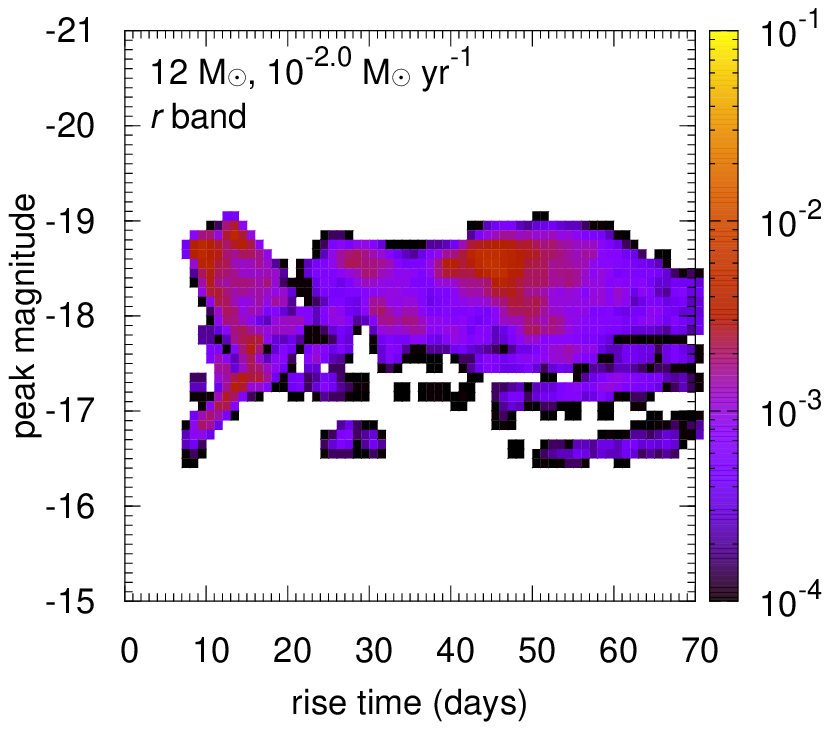}
	\includegraphics[width=0.66\columnwidth]{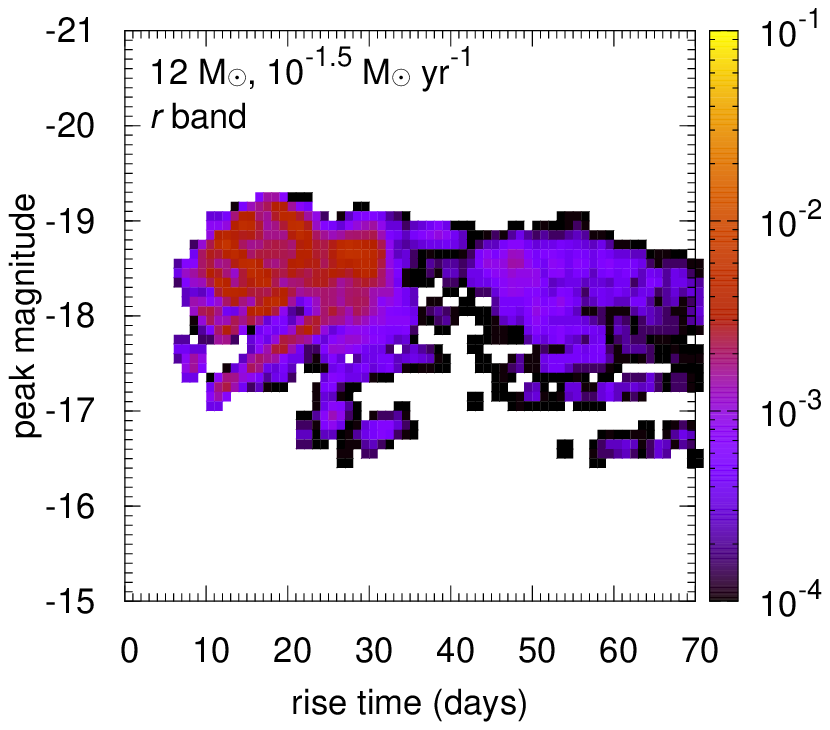}
	\includegraphics[width=0.66\columnwidth]{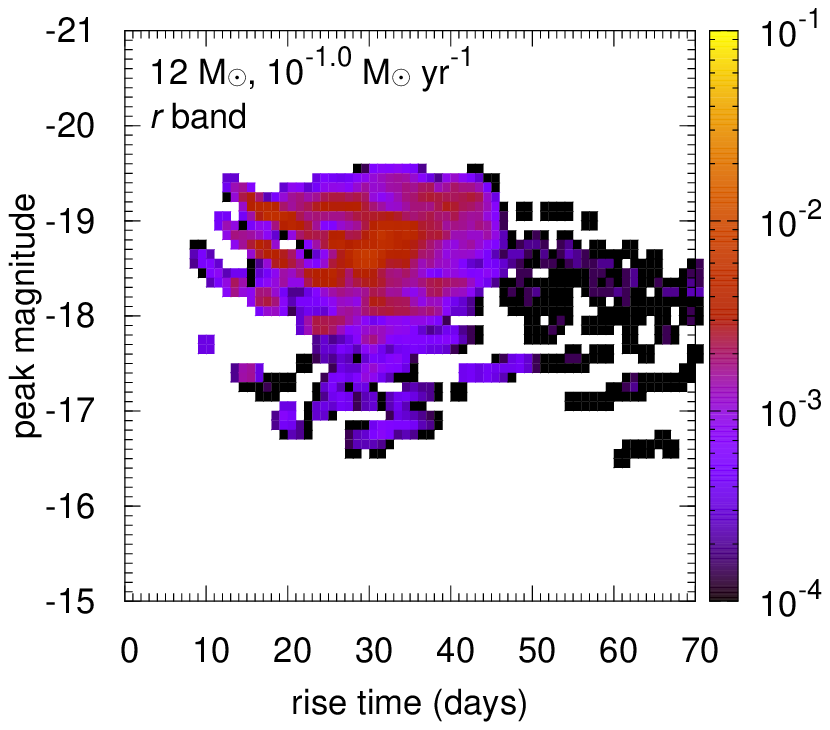}
  \end{center}
  \caption{%
  Same as Fig.~\ref{fig:s12_risepeak_g_mdot}, but for the LSST \textit{r} band filter.
}%
  \label{fig:s12_risepeak_r_mdot}
\end{figure*}

The short rise times observed in Type~II SNe (e.g., \cite{gonzalez-gaitan2015,gall2015}) have been related to the existence of dense confined CSM around Type~II SN progenitors (e.g., \cite{forster2018}). We discuss how the rise time and peak luminosity in Type~II SNe are affected by confined CSM. Figures~\ref{fig:s12_risepeak_g_mdot} and \ref{fig:s12_risepeak_r_mdot} present how mass-loss rates affect the rise time and peak luminosity distributions in the LSST \textit{g} and \textit{r} band, respectively, for the case of the $12~\Msun$ progenitor. The \textit{g} band rise time is around 20~days when the mass-loss rate is small. The peak magnitude gap at around $-16.5~\mathrm{mag}$ is caused by the lack of models between 0.5~B and 1.0~B, i.e., most 0.5~B explosion models have peak magnitudes fainter than $-16.5~\mathrm{mag}$ while most models about 1.0~B have peak magnitudes brigther than $-16.5~\mathrm{mag}$. We start to find a population of Type~II SNe with short rise times of around 5~days when $\dot{M}\gtrsim 10^{-3.0}~\Msunpyr$. As the mass-loss rate increases, the diffusion time in CSM becomes longer and thus the rise time becomes longer again at $\dot{M}\gtrsim 10^{-1.5}~\Msunpyr$. The \textit{r} band rise time is less sensitive to mass-loss rates, but we start to see the similar effects in the \textit{r} band rise time when $\dot{M}\gtrsim 10^{-2.5}~\Msunpyr$.

\begin{figure*}
  \begin{center}
	\includegraphics[width=0.66\columnwidth]{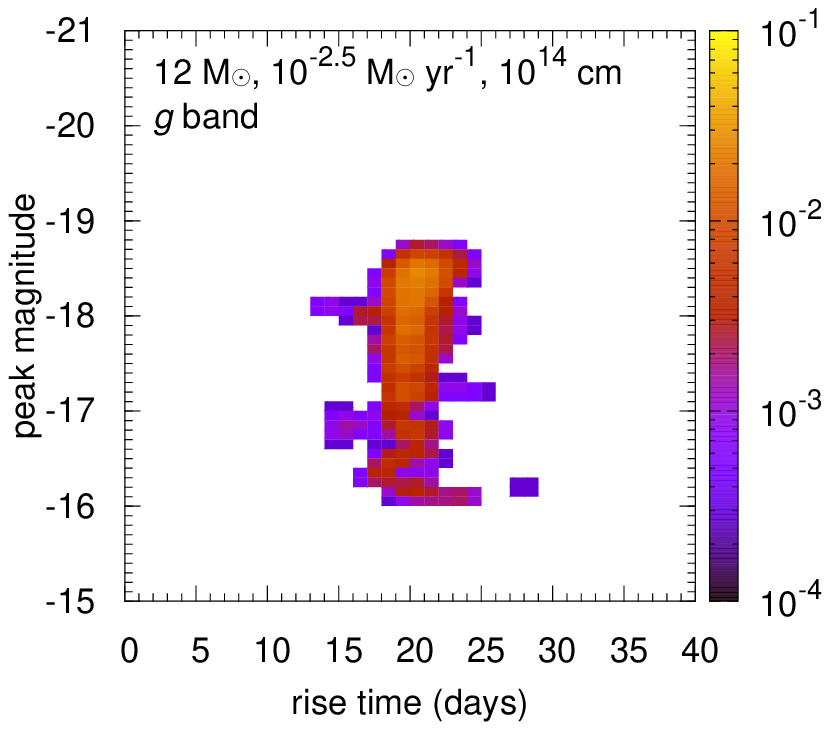}
	\includegraphics[width=0.66\columnwidth]{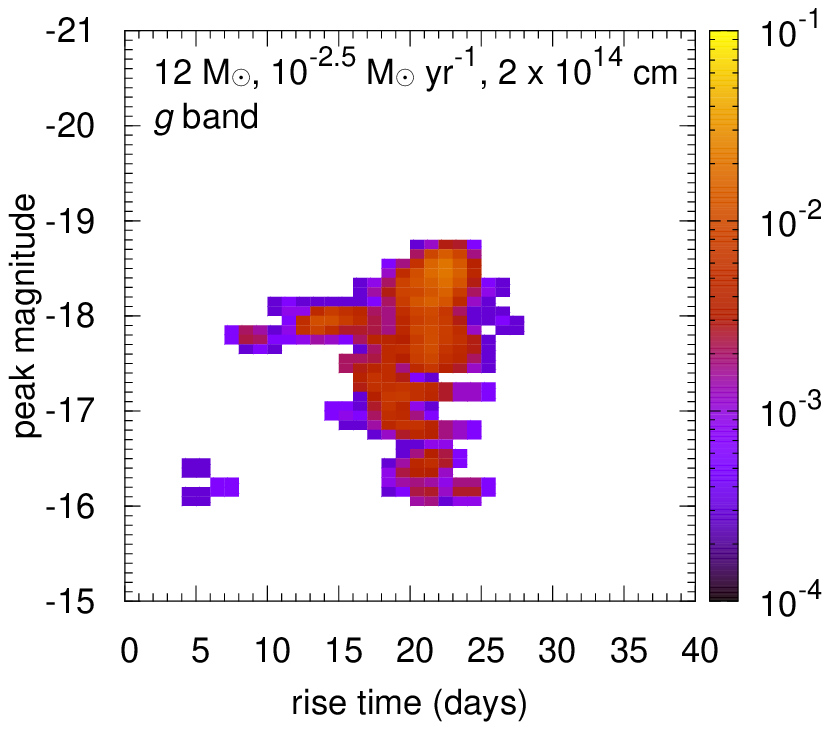}
	\includegraphics[width=0.66\columnwidth]{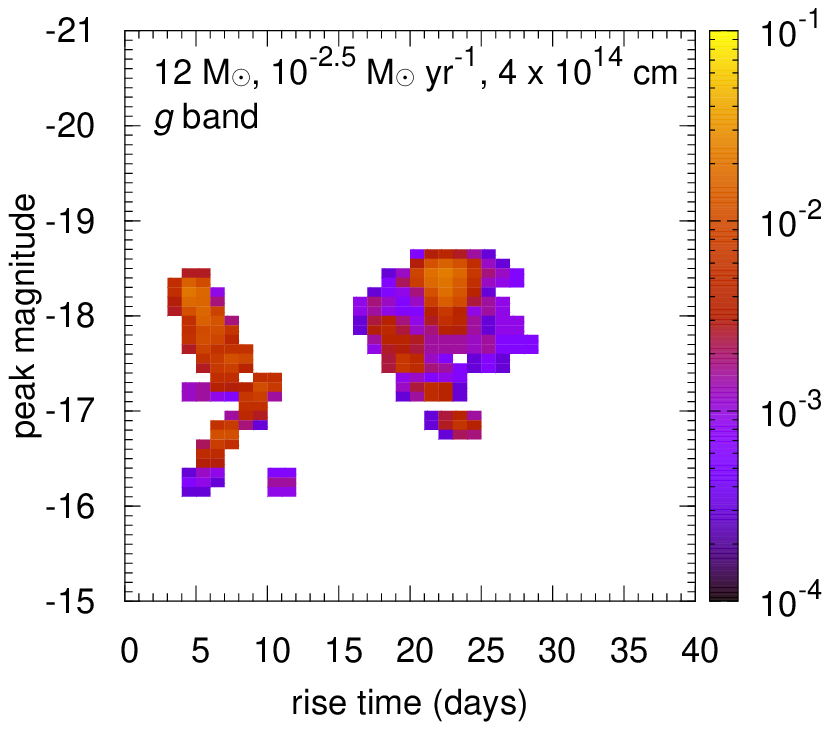}
	\includegraphics[width=0.66\columnwidth]{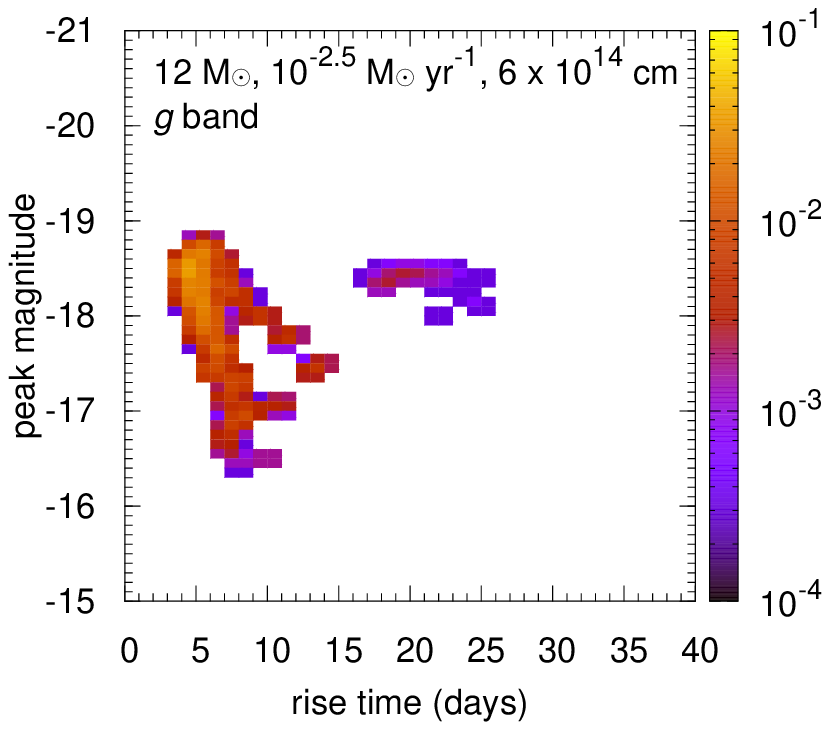}
	\includegraphics[width=0.66\columnwidth]{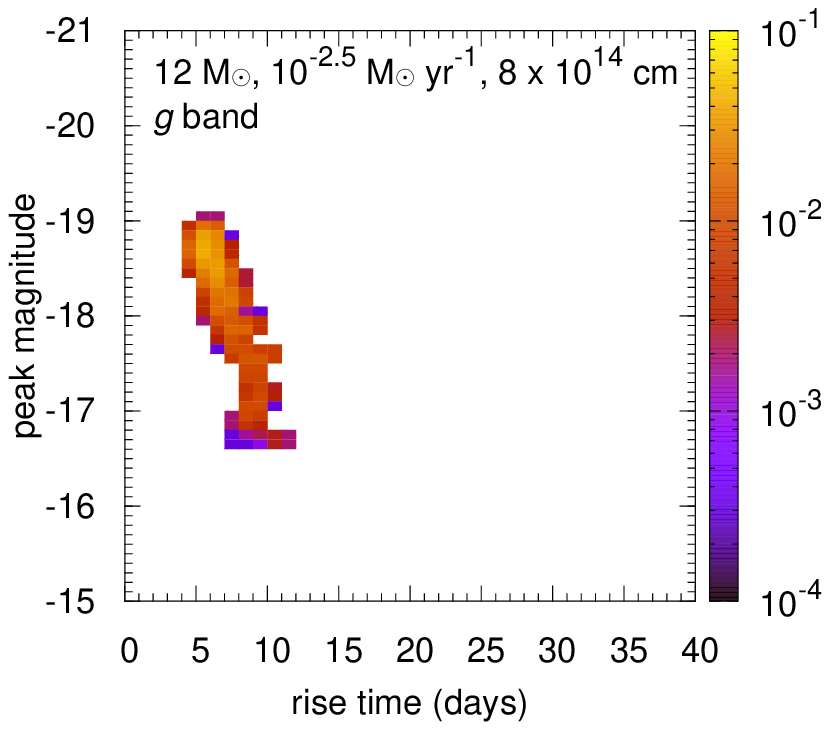}
	\includegraphics[width=0.66\columnwidth]{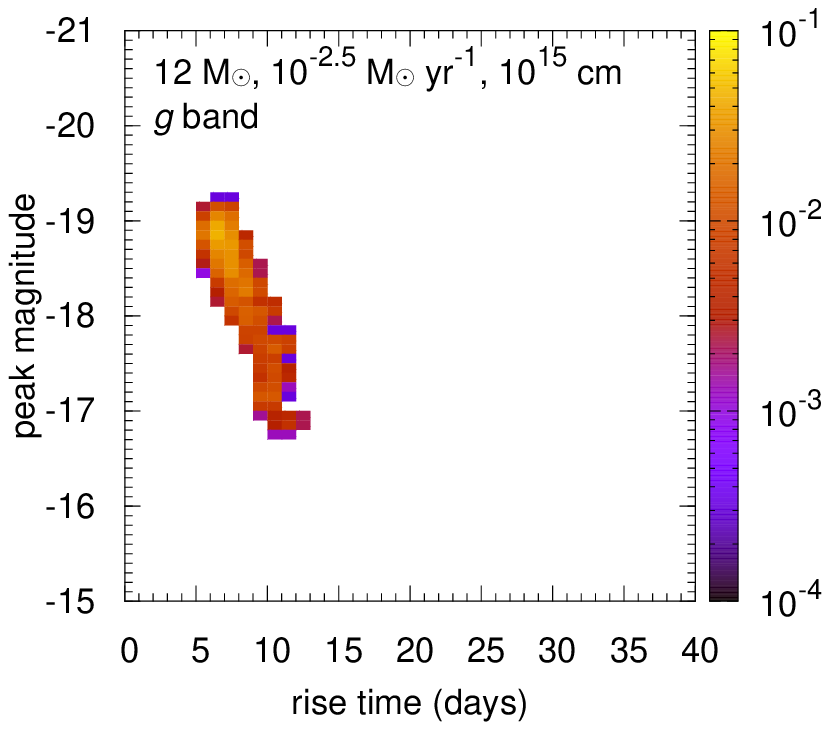}
  \end{center}
  \caption{%
  Rise time and peak luminosity distributions of Type~II SNe from the 12~\Msun\ progenitor in the LSST \textit{g} band with the same mass-loss rate ($10^{-2.5}~\Msunpyr$), but with different CSM radii. The color contours show relative fractions of the models within each bin. Each panel shows a summary of all the models with the CSM radius presented in the panel.
}%
  \label{fig:s12_risepeak_g_radius}
\end{figure*}

When $\dot{M}\gtrsim 10^{-3.0}~\Msunpyr$, we can still find some populations of Type~II SNe with long rise times as in the lower $\dot{M}$ models in Fig.~\ref{fig:s12_risepeak_g_mdot}. This is because of the existence of CSM with small radii. Figure~\ref{fig:s12_risepeak_g_radius} shows how the CSM radius affects the rise time and peak luminosity in the LSST \textit{g} band for a fixed mass-loss rate of $10^{-2.5}~\Msunpyr$. Even if the mass-loss rate is high, a population of short rise times remain if a CSM radius is small ($\lesssim 2\times 10^{14}~\mathrm{cm}$). The population of Type~II SNe with short rise times start to appear when the CSM radius is more than $4\times 10^{14}~\mathrm{cm}$.

\begin{figure*}
  \begin{center}
	\includegraphics[width=0.66\columnwidth]{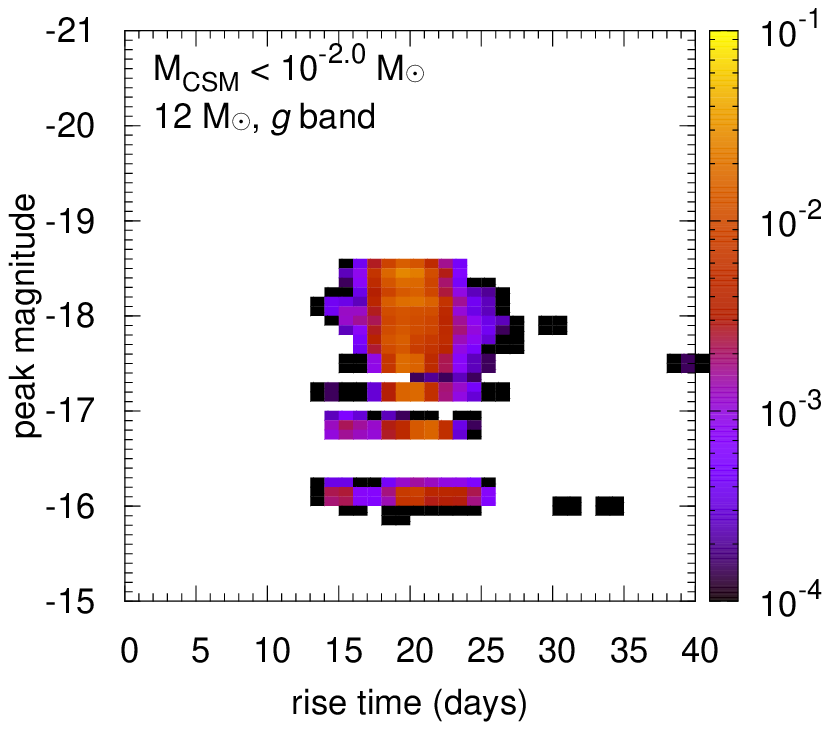}
	\includegraphics[width=0.66\columnwidth]{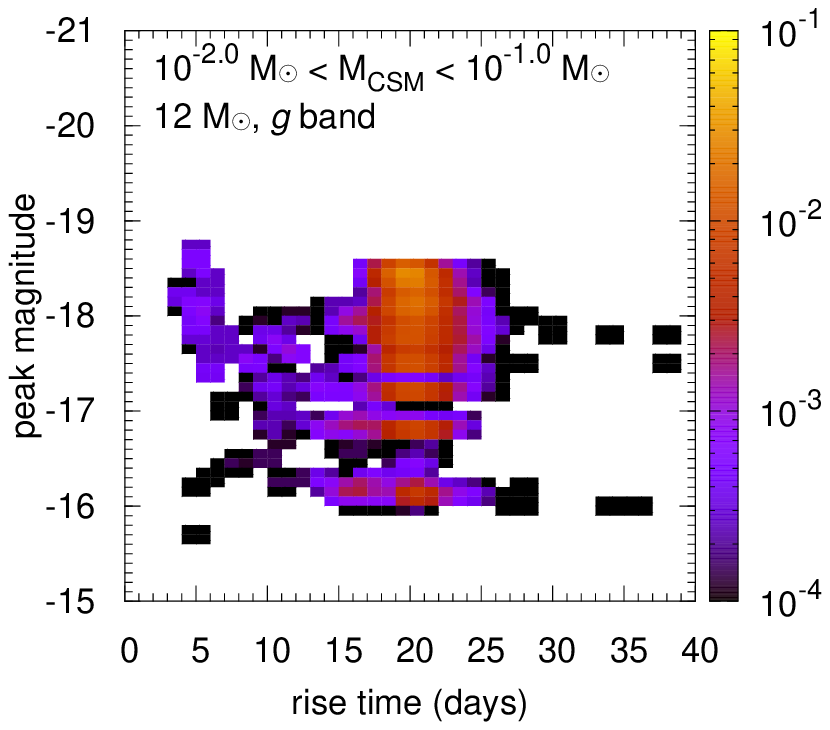}
	\includegraphics[width=0.66\columnwidth]{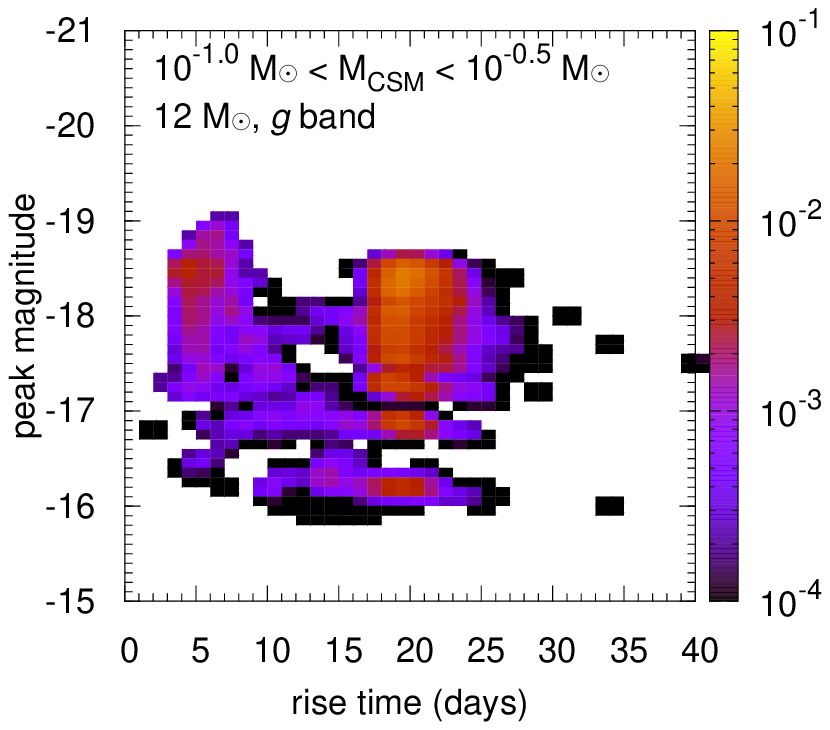}
	\includegraphics[width=0.66\columnwidth]{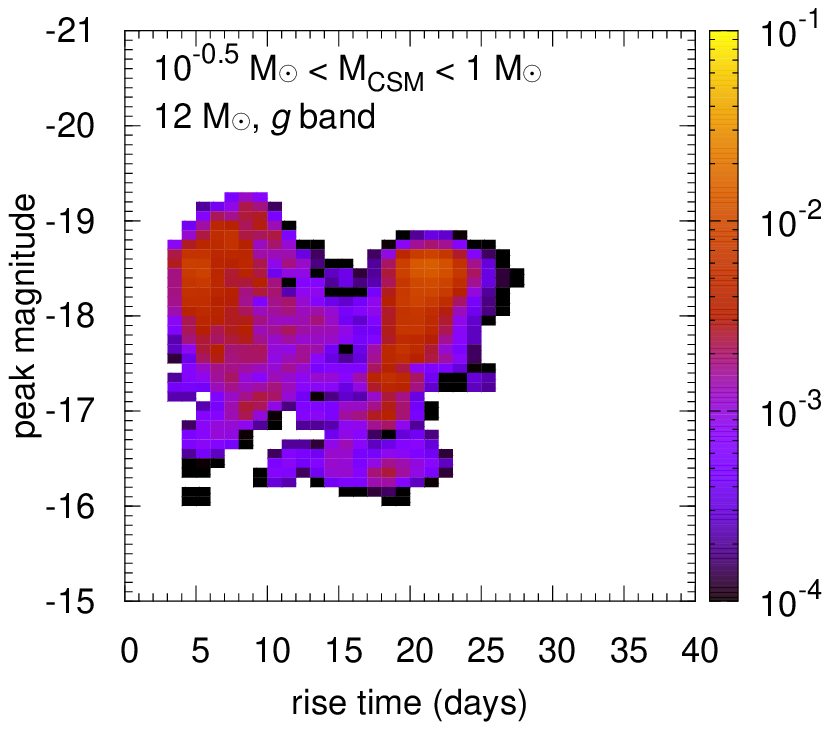}
	\includegraphics[width=0.66\columnwidth]{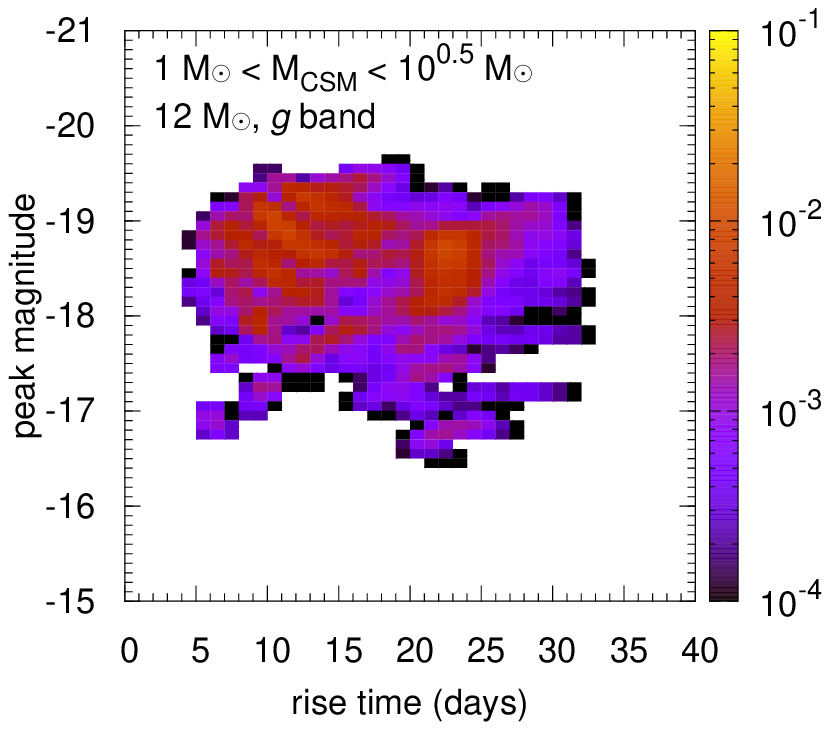}
	\includegraphics[width=0.66\columnwidth]{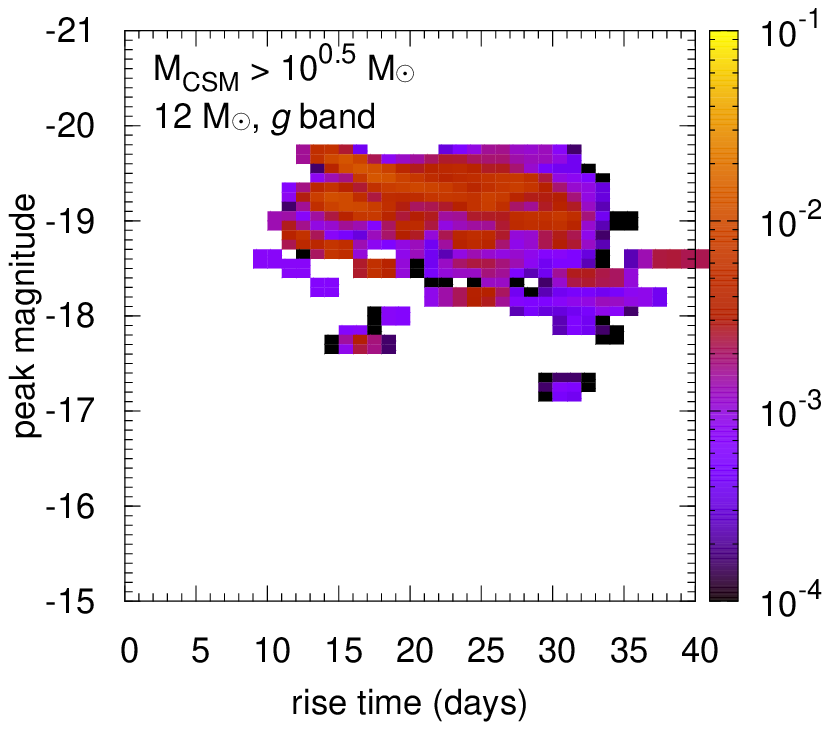}
  \end{center}
  \caption{%
  Effects of CSM mass in the rise time and peak luminosity distributions of Type~II SNe from the 12~\Msun\ progenitor in the LSST \textit{g} band. The color contours show relative fractions of the models within each bin. Each panel shows a summary of all the models with the same CSM mass.
}%
  \label{fig:s12_risepeak_g_csm}
\end{figure*}

Fig.~\ref{fig:s12_risepeak_g_csm} demonstrates how the CSM mass ($M_\mathrm{CSM}$) affects the rise time and peak luminosity distributions in the LSST \textit{g} band. The population of Type~II SNe with short rise times can be found when $M_\mathrm{CSM}\gtrsim 0.1~\Msun$. When the CSM mass exceeds about $1~\Msun$, the rise time becomes long because of large diffusion time in CSM.

\begin{figure*}
  \begin{center}
	\includegraphics[width=0.66\columnwidth]{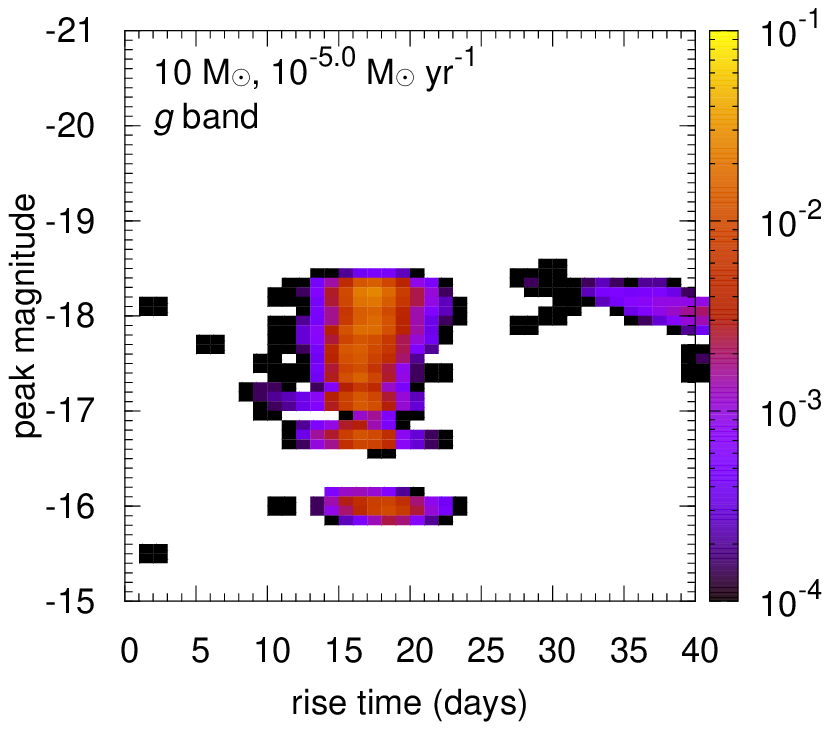}
	\includegraphics[width=0.66\columnwidth]{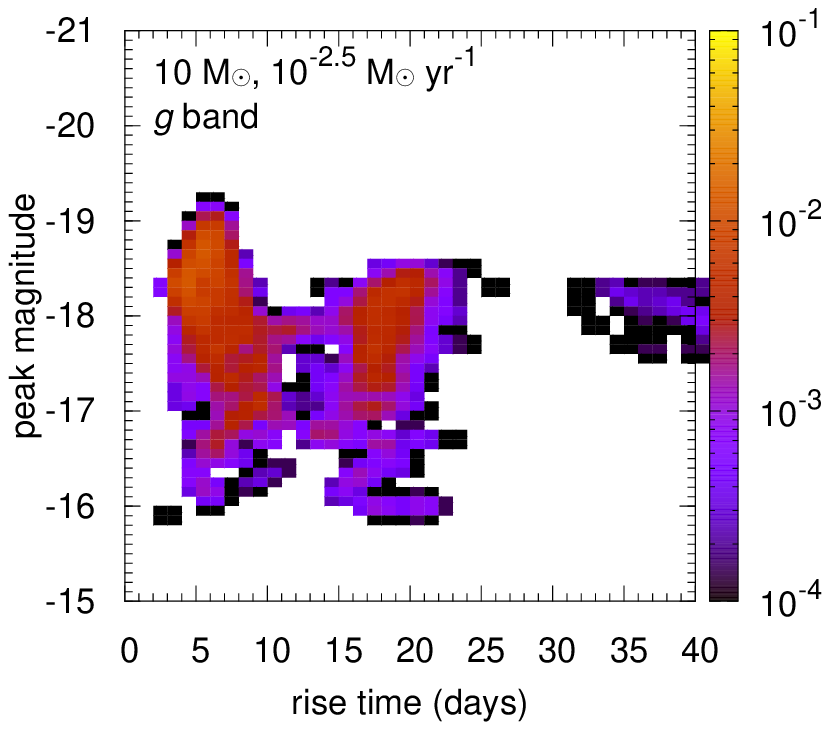}
	\includegraphics[width=0.66\columnwidth]{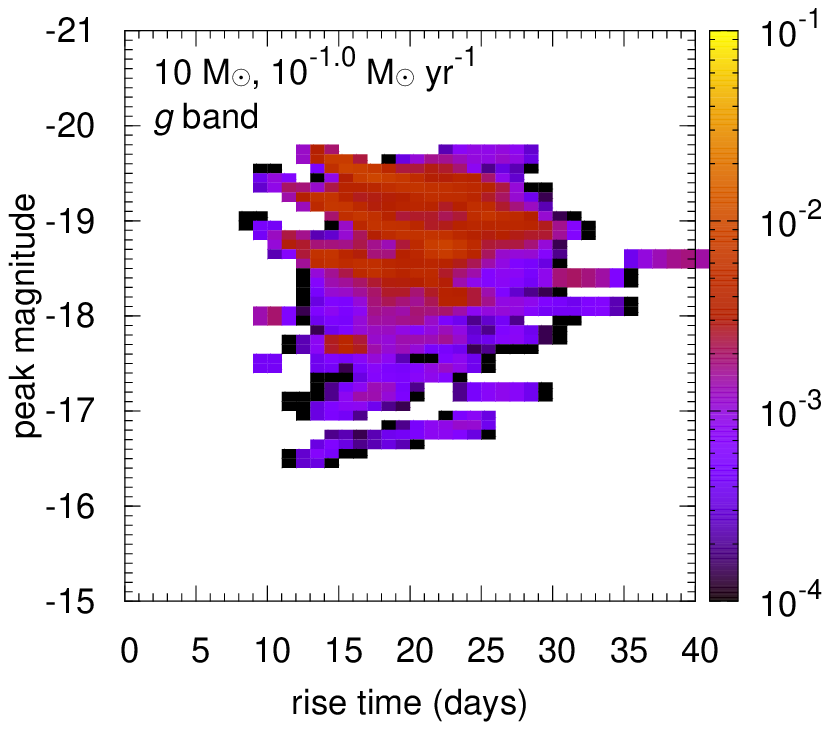}
	\includegraphics[width=0.66\columnwidth]{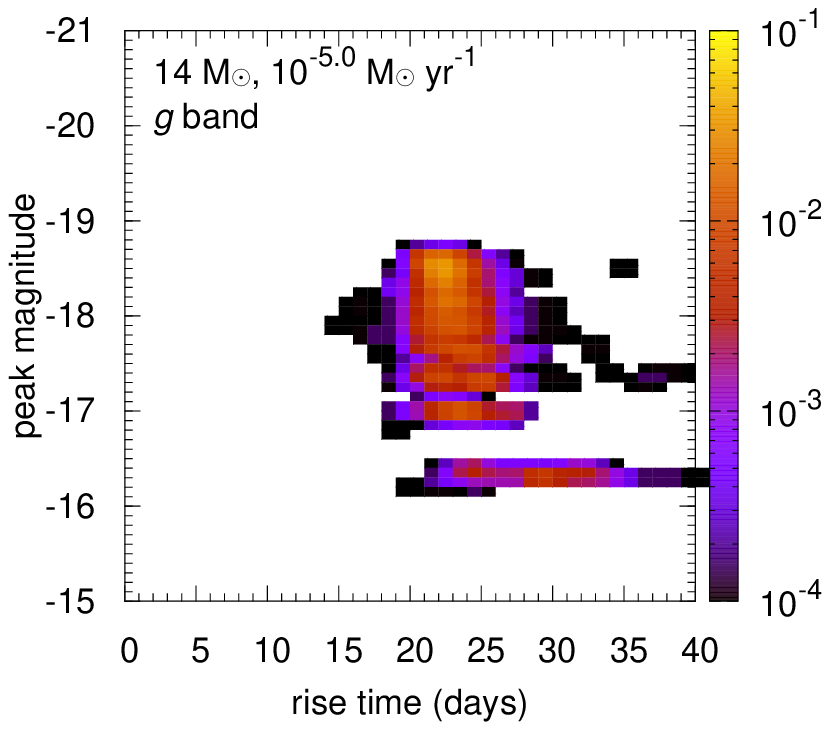}
	\includegraphics[width=0.66\columnwidth]{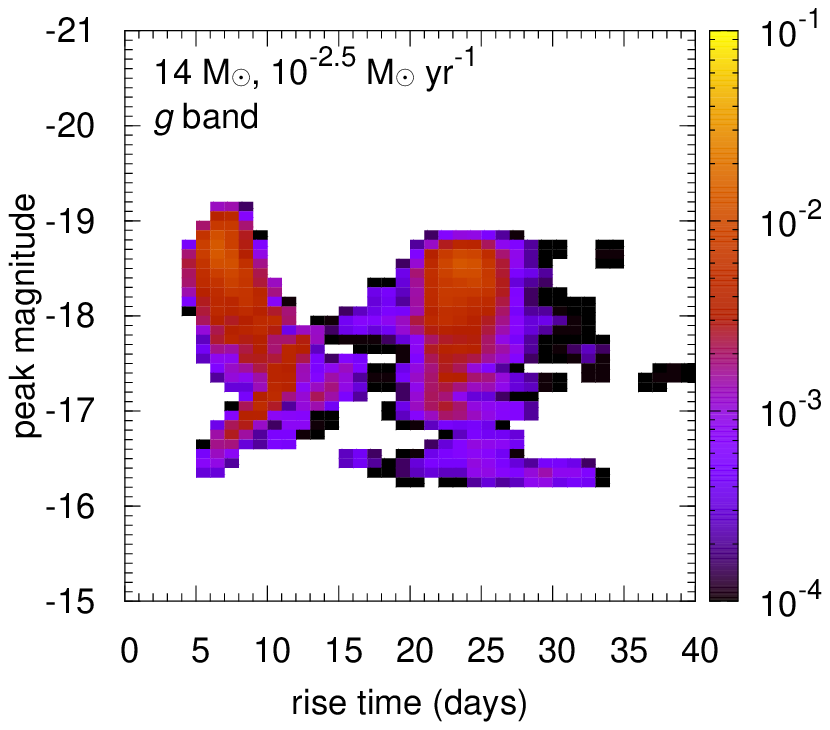}
	\includegraphics[width=0.66\columnwidth]{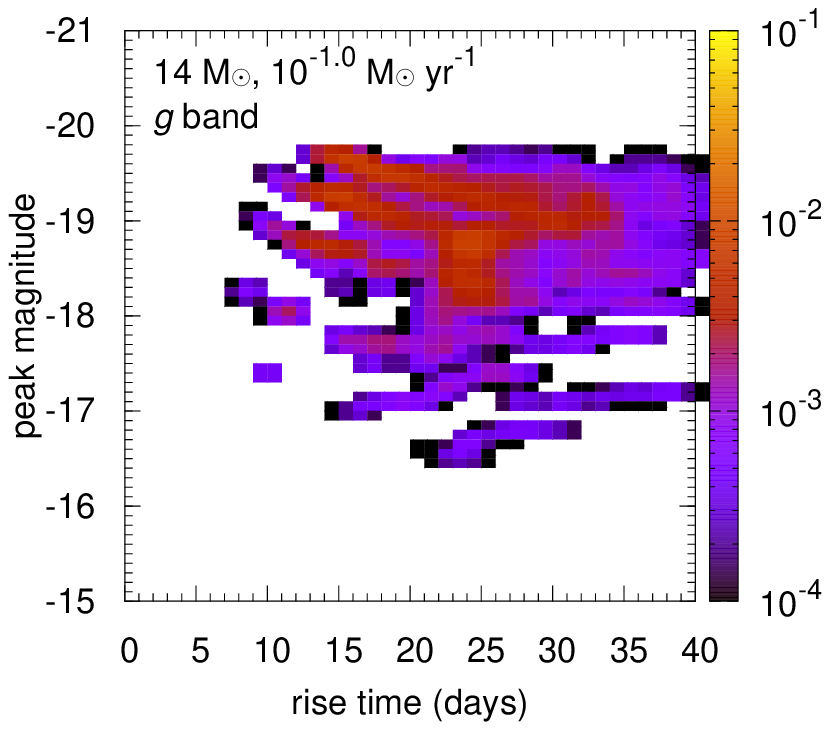}
	\includegraphics[width=0.66\columnwidth]{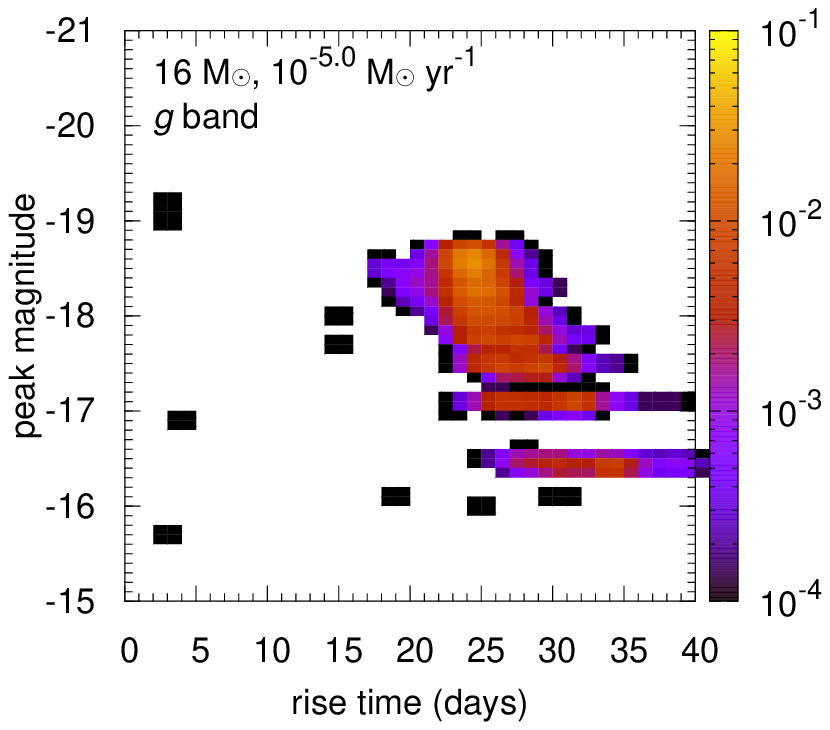}
	\includegraphics[width=0.66\columnwidth]{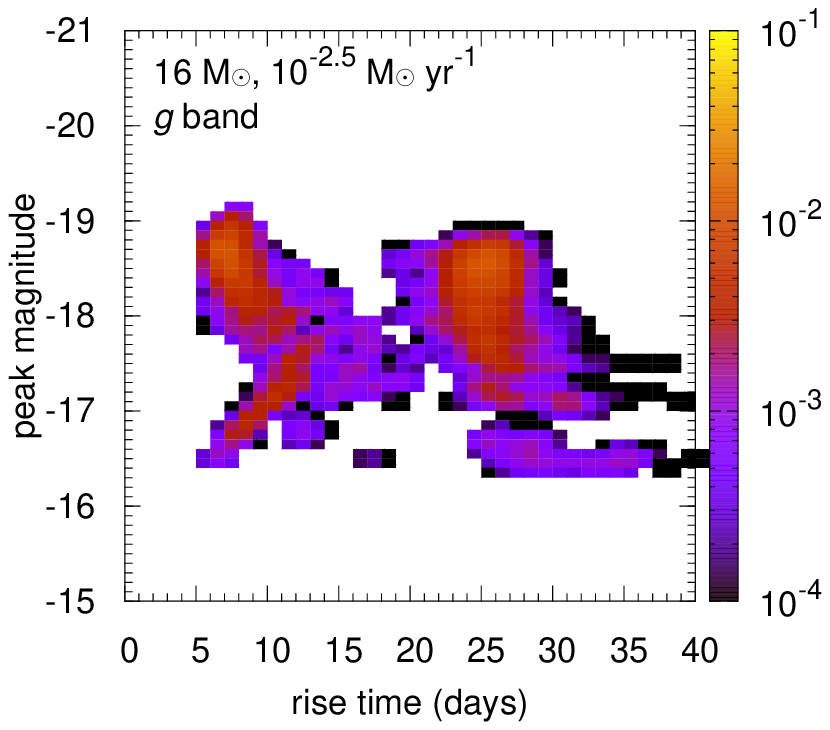}
	\includegraphics[width=0.66\columnwidth]{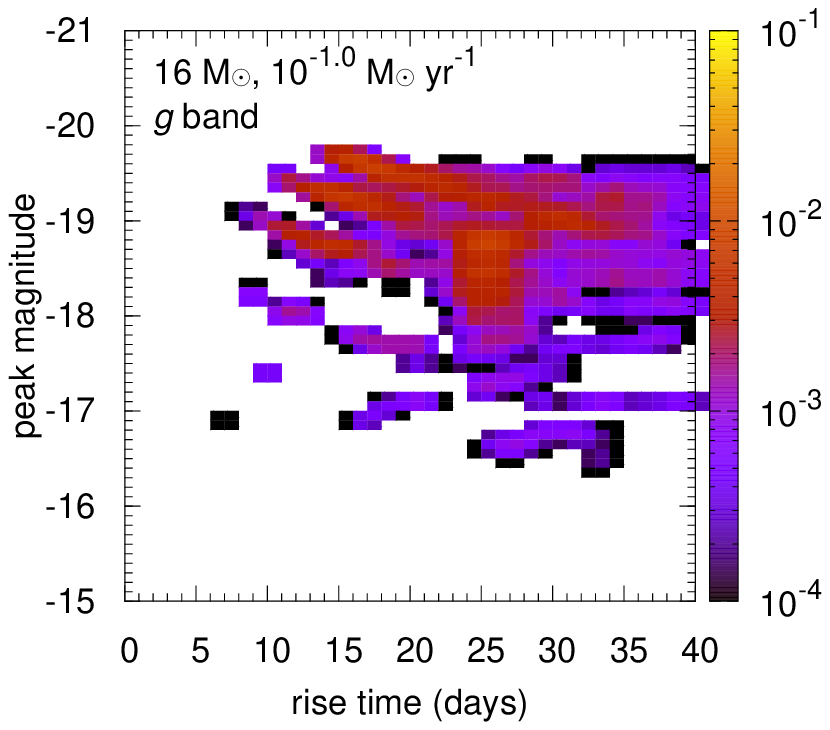}
	\includegraphics[width=0.66\columnwidth]{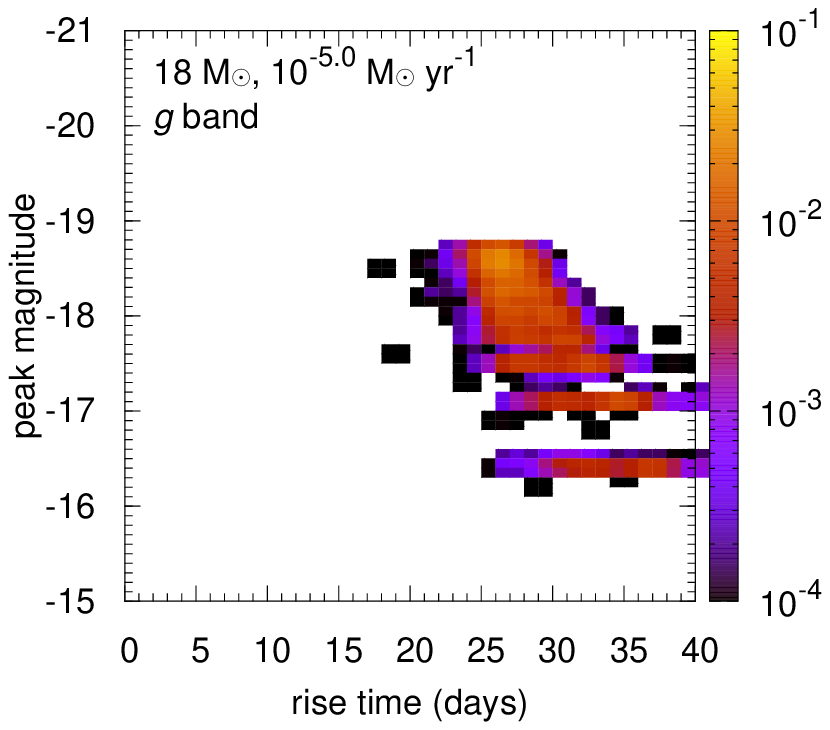}
	\includegraphics[width=0.66\columnwidth]{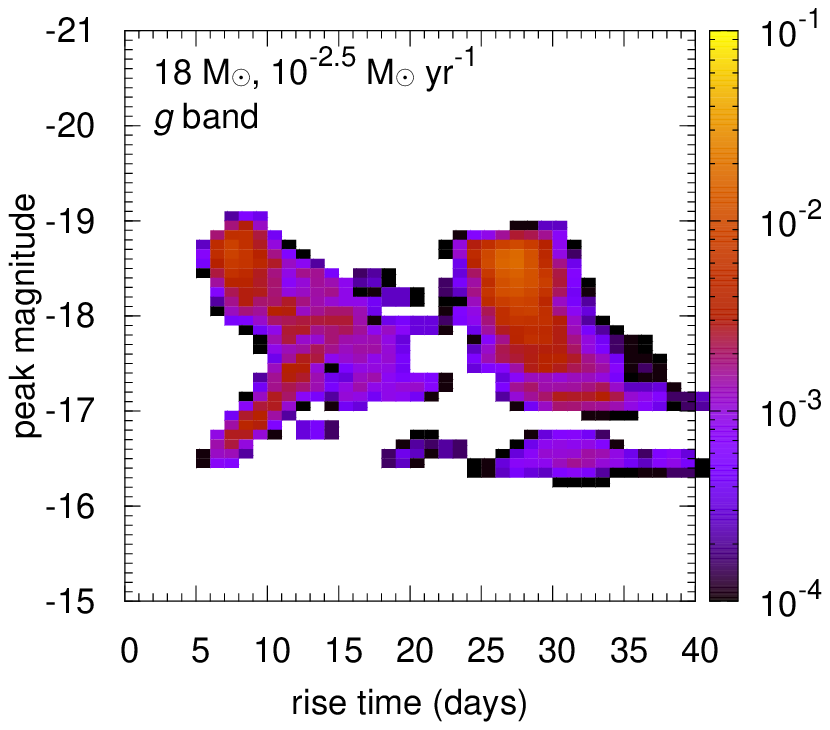}
	\includegraphics[width=0.66\columnwidth]{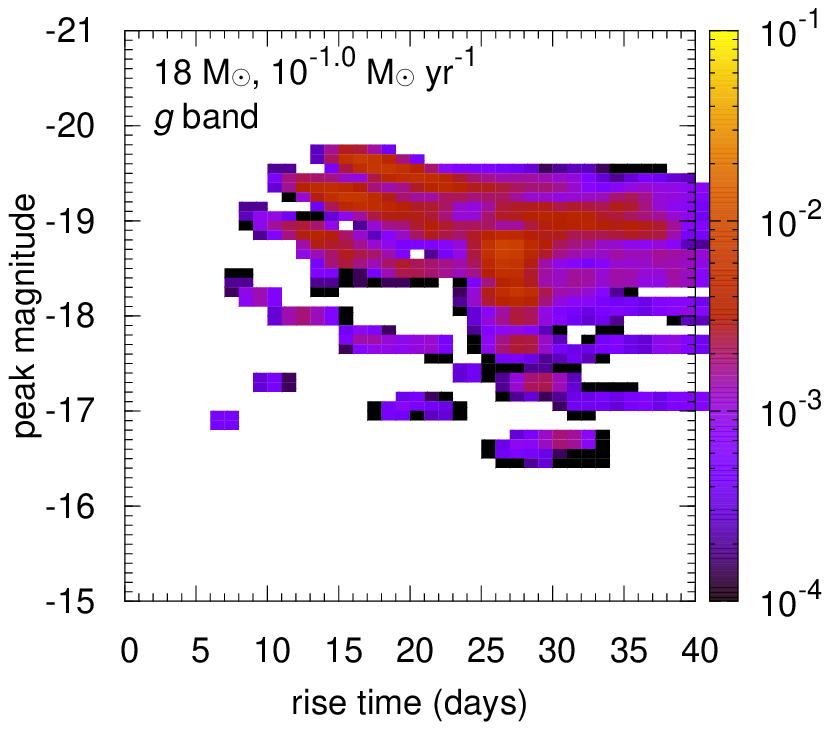}
  \end{center}
  \caption{%
  Rise time and peak luminosity distributions of Type~II SNe in the LSST \textit{g} band from the $10,14,16,$ and $18~\Msun$ progenitors. The color contours show relative fractions of the models within each bin. Each panel shows a summary of all the models with one mass-loss rate.
}%
  \label{fig:progenitor_risepeak_g}
\end{figure*}

We focused on the $12~\Msun$ progenitor models so far. Figure~\ref{fig:progenitor_risepeak_g} summarizes the rise time and peak luminosity distributions from the other progenitors in the LSST \textit{g} band. While the rise time and peak luminosity distributions are affected by the differences in the progenitors when the mass-loss rate is low, their dependence on the progenitors becomes less significant as the mass-loss rate becomes higher. This is because they start to be mainly affected by the CSM properties.

\begin{figure}
  \begin{center}
	\includegraphics[width=\columnwidth]{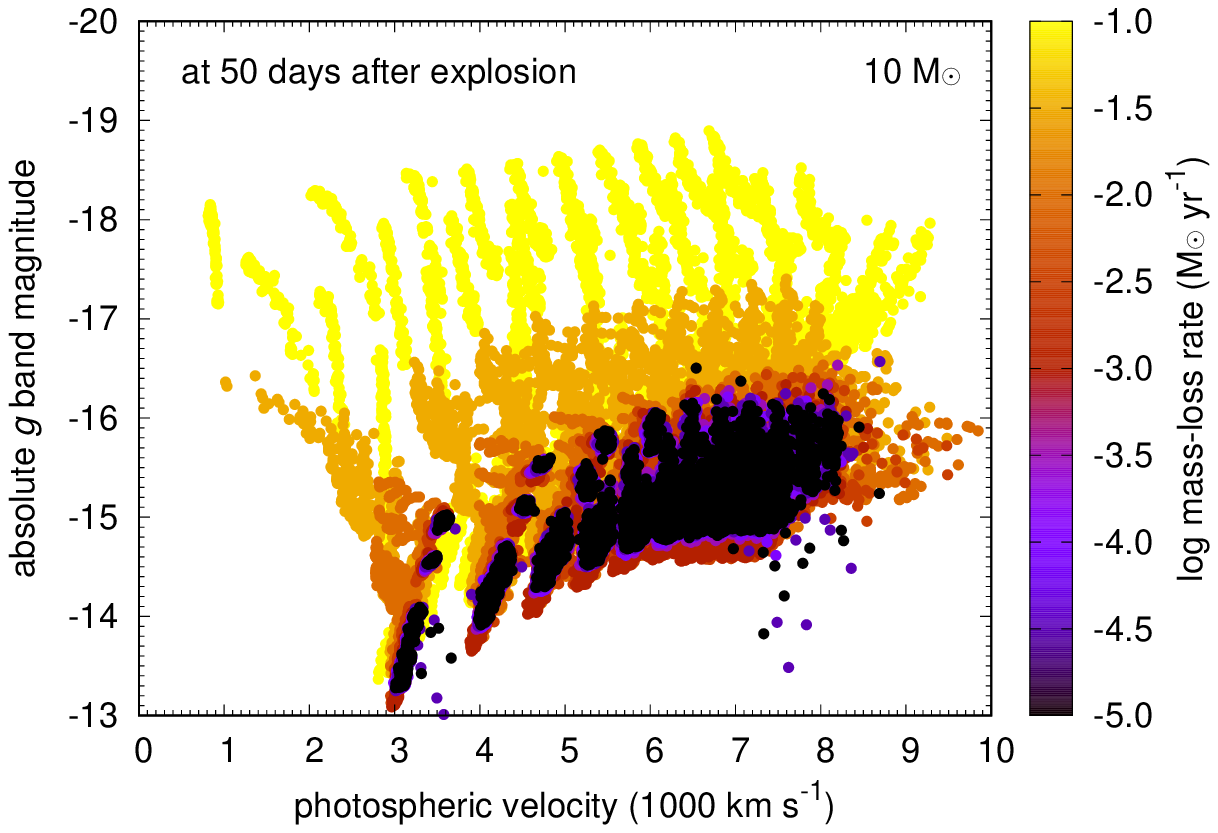}
	\includegraphics[width=\columnwidth]{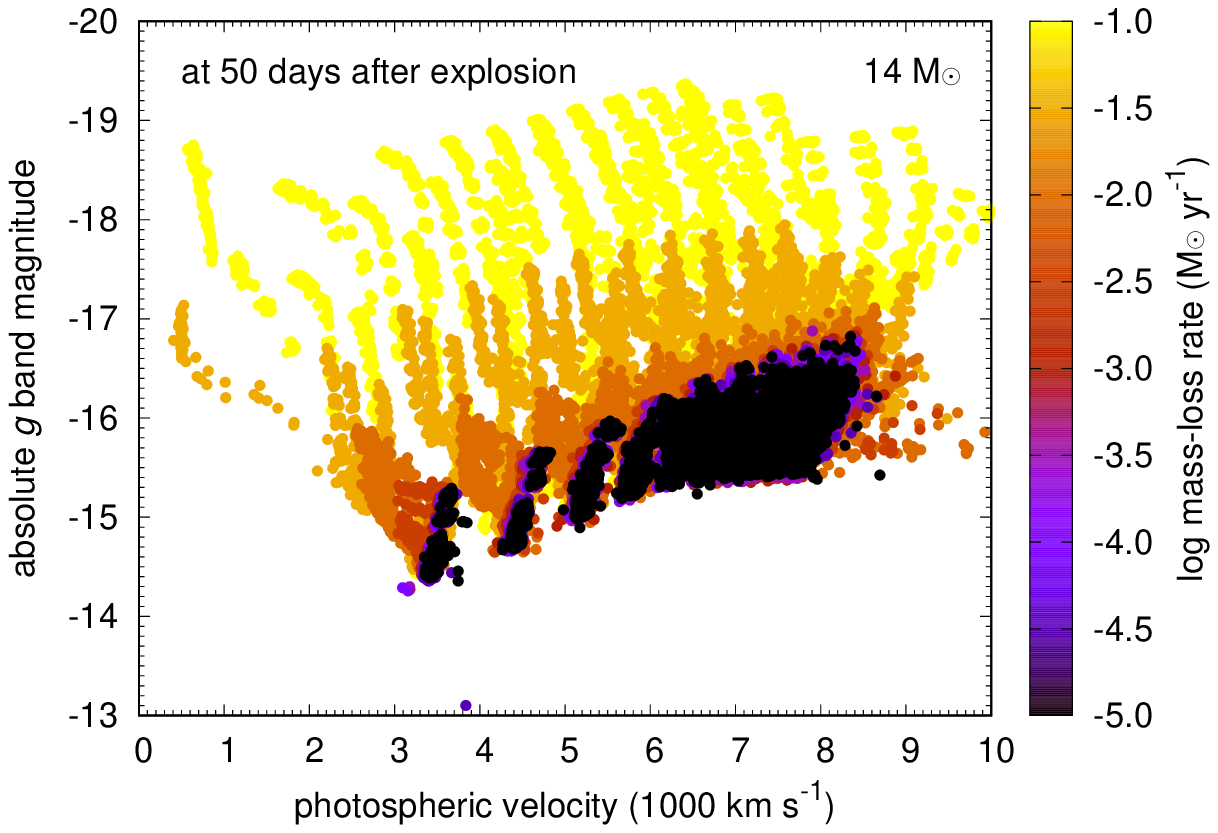}
	\includegraphics[width=\columnwidth]{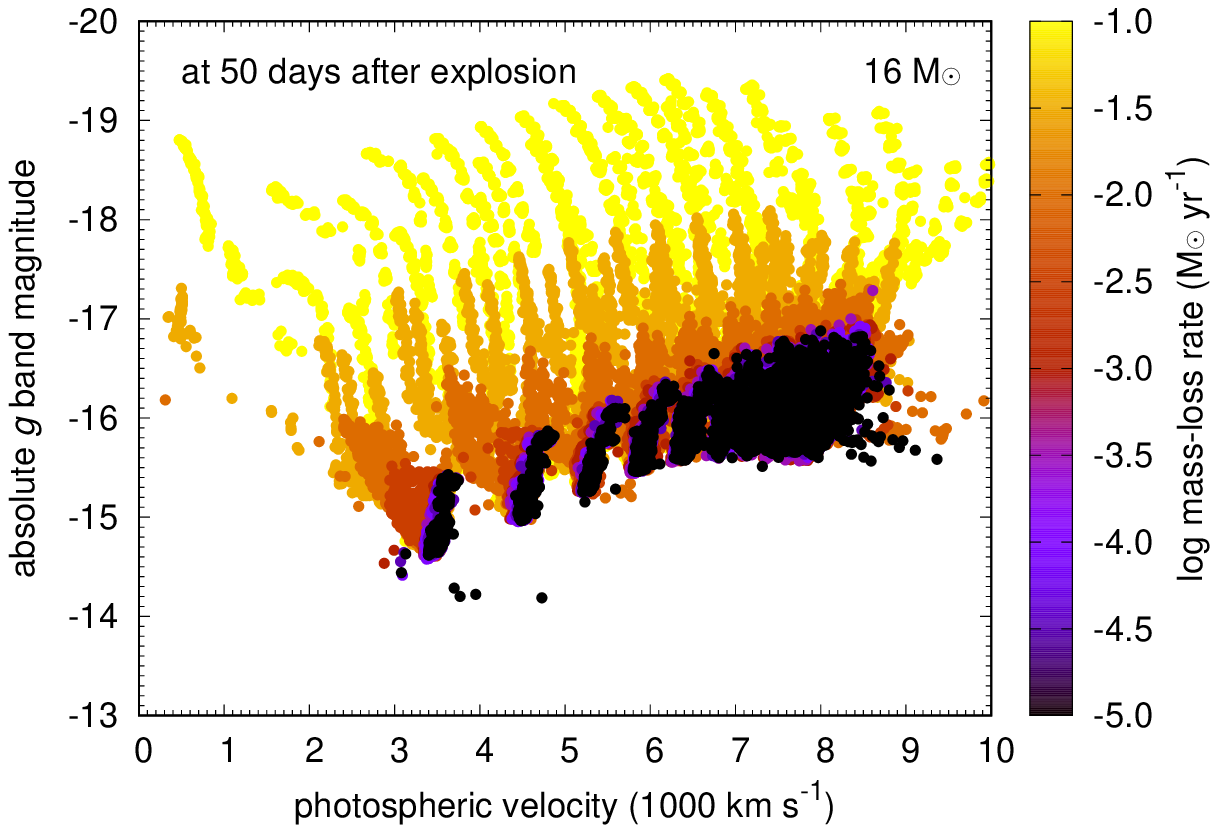}
	\includegraphics[width=\columnwidth]{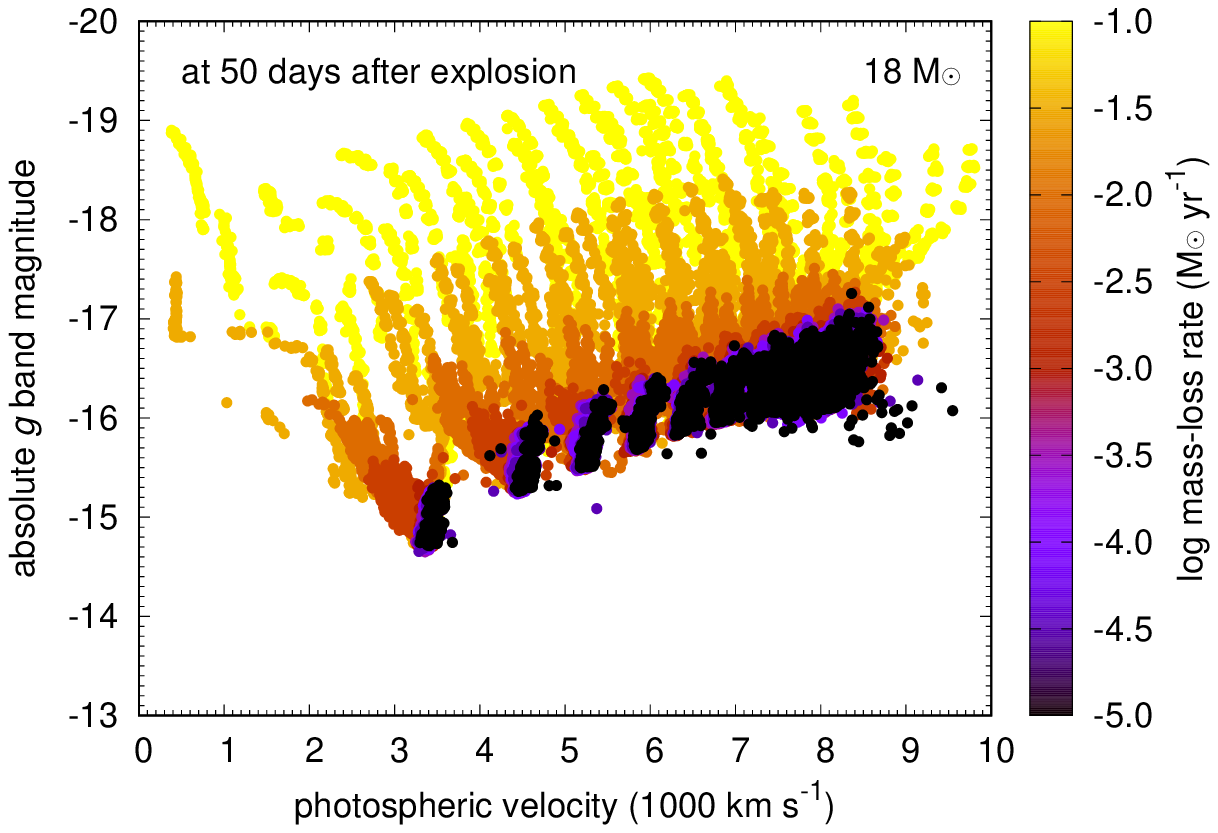}
  \end{center}
  \caption{%
  Relation between photospheric velocity and LSST \textit{g} band magnitude at 50~days after explosion in synthetic models with the $10,14,16,$ and $18~\Msun$ progenitors. The mass-loss rate of each model is shown by color.
}%
  \label{fig:many_velocity}
\end{figure}

Properties of photosphere are provided in the model grid, and they also give important and interesting insight to Type~II SNe. As an example, we present a relation between photospheric velocity and the LSST \textit{g} band magnitude at 50~days after explosion in the $12~\Msun$ progenitor in Fig.~\ref{fig:s12_velocity}. Overall, we can see a trend that explosions with higher explosion energy have higher photospheric velocities and higher luminosities. The relation between photospheric velocity and \textit{g} band magnitude is strongly affected by CSM properties as well. We can find that a population of Type~II SNe with high luminosity is caused by the models with large mass-loss rates exceeding $10^{-2.0}~\Msunpyr$. This bright population appears even if explosion energy and photospheric velocity are small regardless of the progenitor mass (Fig.~\ref{fig:many_velocity}). Such a population is starting to be identified observationally \citep{rodriguez2020}, and it also demonstrates the importance of the role of CSM in understanding the diversity in Type~II SNe.

\section{Summary}\label{sec:summary}
We have introduced a new model grid of Type~II SNe containing 228,016 synthetic explosion models. Parameters that changed are progenitor mass, explosion energy, \Ni\ mass, mass-loss rate, CSM density structure and CSM radius. For each model, we provide SED ($1-50,000$~\AA), absolute magnitudes in the LSST, ZTF, SDSS, and Swift filters in each time step. Photosphere information is also provided. Light curves in any photometric bands can be constructed by using SEDs. Our model grid confirms the importance of confined CSM in shaping the diversities observed in Type~II SNe. The model grid is available at \url{https://doi.org/10.5061/dryad.pnvx0k6sj}.

In the upcoming era of LSST, rapid characterization of Type~II SNe with limited photometry information will be required. Thus, our synthetic model grid of Type~II SNe that allows direct comparison of observations and models is expected to be important. The new model grid can be a base for a Bayesian parameter estimation of Type~II SNe that has already been demonstrated to be powerful \citep{subrayan2022}.

\begin{figure}
  \begin{center}
	\includegraphics[width=\columnwidth]{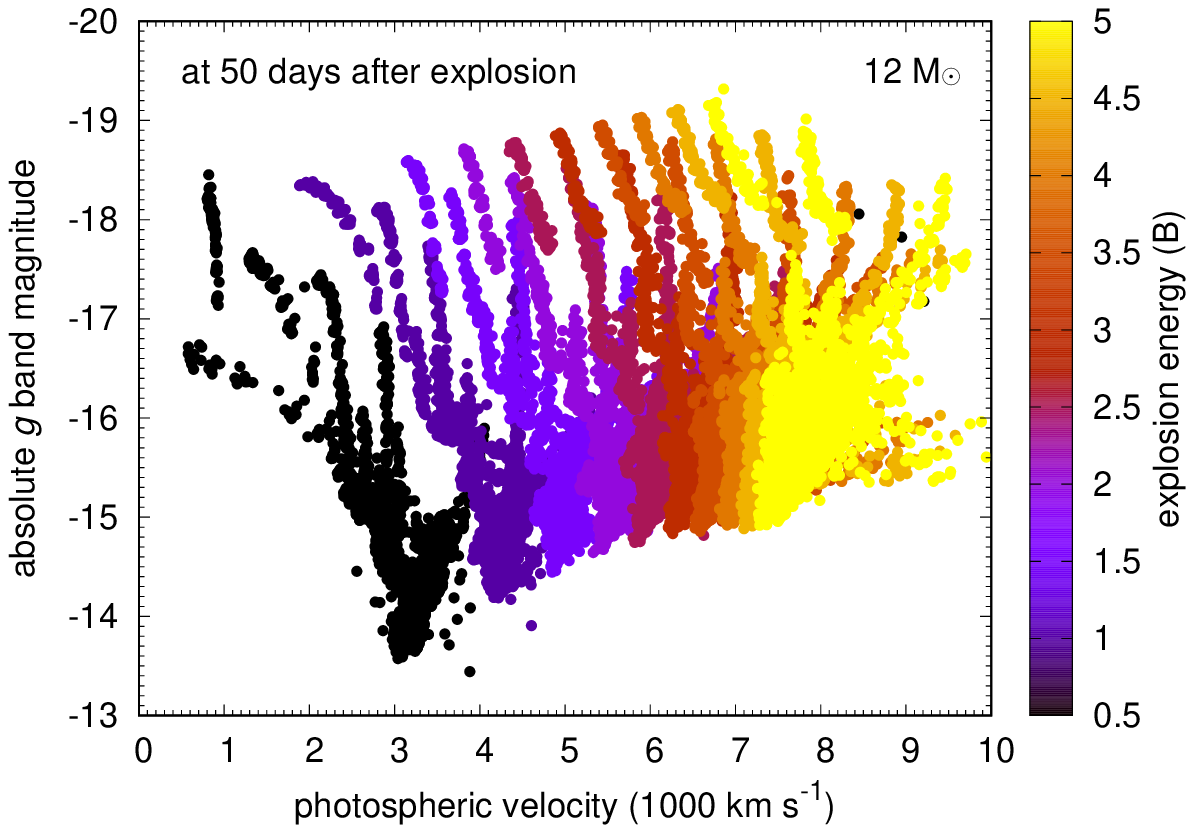}
	\includegraphics[width=\columnwidth]{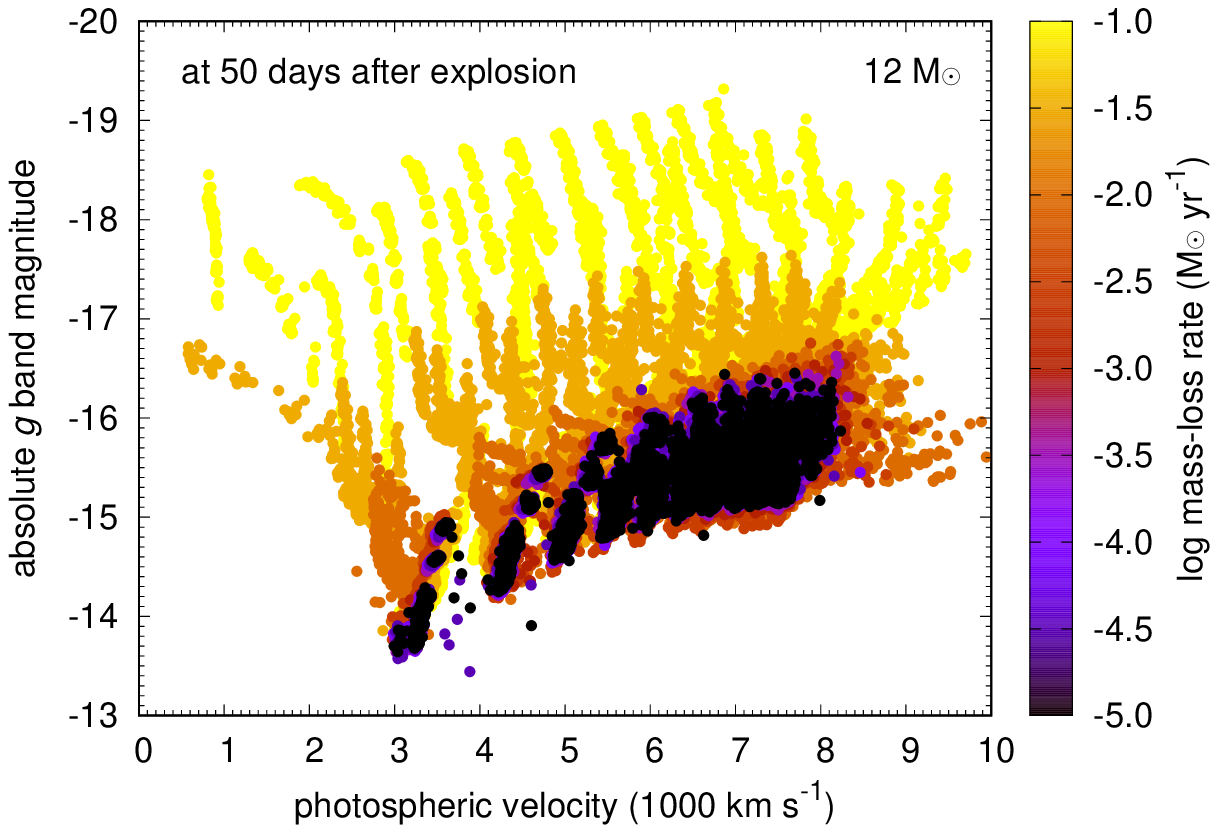}
  \end{center}
  \caption{%
  Relation between photospheric velocity and LSST \textit{g} band magnitude at 50~days after explosion in synthetic models with the $12~\Msun$ progenitor. The top panel presents the explosion energy of each model while the bottom panel shows the mass-loss rate of each model.
}%
  \label{fig:s12_velocity}
\end{figure}

This model set can also be used to search for predicted properties of Type~II SNe. For example, bright Type~II SNe with low photospheric velocity are predicted to be mostly caused by Type~II SNe with high mass-loss rates exceeding $10^{-2.0}~\Msunpyr$ (Fig.~\ref{fig:s12_velocity}). More investigation into the model grid properties would also help us understand the origin of diversity in Type~II SNe.

\begin{ack}
Numerical computations were carried out on PC cluster at Center for Computational Astrophysics (CfCA), National Astronomical Observatory of Japan.
TJM is supported by the Grants-in-Aid for Scientific Research of the Japan Society for the Promotion of Science (JP20H00174, JP21K13966, JP21H04997).
SIB is supported by RSCF grant 19-12-00229 in his work on SN simulations.
This research was partly supported by the Munich Institute for Astro-, Particle and BioPhysics (MIAPbP) which is funded by the Deutsche Forschungsgemeinschaft (DFG, German Research Foundation) under Germany´s Excellence Strategy – EXC-2094 – 390783311.
This research has made use of the Spanish Virtual Observatory (\url{https://svo.cab.inta-csic.es}) project funded by MCIN/AEI/10.13039/501100011033/ through grant PID2020-112949GB-I00 \citep{rodrigo2012,rodrigo2020}.
\end{ack}

\bibliographystyle{apj}
\bibliography{pasj}

\end{document}